\documentclass[amsmath,showpacs,aps,twocolumn,prb]{revtex4-1}

\usepackage{bm}
\usepackage{graphics}
\usepackage{graphicx}
\usepackage[dvips]{epsfig}

\begin{document}
\title{Topological Insulators and Quantum Spin Liquids}
\author{Gregory A. Fiete}
\author{Victor Chua}
\author{Mehdi Kargarian}
\author{Rex Lundgren}
\author{Andreas R\"uegg}
\author{Jun Wen}
\author{Vladimir Zyuzin}
\affiliation{Department of Physics, The University of Texas at Austin, Austin, TX 78712, USA}
\begin{abstract}
In this paper we review some  connections recently discovered between topological insulators and certain classes of quantum spin liquids, focusing on two and three spatial dimensions.  In two dimensions we show the integer quantum Hall effect plays a key role in relating topological insulators and chiral spin liquids described by fermionic excitations, and we describe a procedure for ``generating" a certain class of topological states.  In three dimensions we discuss interesting relationships between certain quantum spin liquids and interacting ``exotic" variants of topological insulators.  We focus attention on better understanding interactions in topological insulators, and the phases nearby in parameter space that might result from moderate to strong interactions in the presence of strong spin-orbit coupling.  We stress that oxides with heavy transition metal ions, which often host a competition between electron interactions and spin-orbit coupling, are an excellent place to search for unusual topological phenomena and other unconventional phases.  
\end{abstract}
\date{\today}


\maketitle

\global\long\def\Ju{J_{\triangle}}
\global\long\def\Jd{J_{\nabla}}
\global\long\def\ij{\langle ij\rangle}

\section{Introduction}

At first glance topological insulators and quantum spin liquids have nothing to do with each other.  A topological insulator (TI) has properties that can be understood in the absence of any electron interactions,\cite{Qi:pt10,Moore:nat10,Hasan:rmp10} while quantum spin liquids (QSL) are largely understood theoretically in terms of lattice models of interacting spins.\cite{Balents:nat10}  Moreover, topological insulators explicitly possess charge and spin degrees of freedom, while spin liquids are understood only in terms of spin degrees of freedom.  Quantum spin liquids come in both gapped and gapless varieties, but even for the gapped variety with topological order it would appear there is little to say by way of a relationship between TIs and QSLs.  As a supporting argument, one might point out that the ``type" of topological order is different: Topological insulators do not have any non-trivial ground state degeneracy, while some of the best understood topological spin liquids possess a non-trivial ground state degeneracy.\cite{Wen} 

However, to dismiss any relationship between these two phases of matter is to ignore some beautiful, and we believe insightful, relationships.  To help motivate why it is useful to look for connections between systems that apparently have no relationship to one another we need look no further than Laughlin's account of his discovery of the fractional quantum Hall effect.\cite{Laughlin:rmp99} According to Laughlin, a one-dimensional polymer known as polyacetylene (which displays charge fractionalization, at least theoretically\cite{Su:prl79}) was an important motivation for his inspired guess for a fractional quantum Hall state wavefunction.\cite{Laughlin:rmp99} As it so turns out, there are also deep connections between polyacetylene and topological insulators.\cite{Qi:prb08,Qi:np08}  We believe analogous relationships between TIs and QSLs can also provide useful insights, and could play a key role in deepening our understanding of strongly interacting variants of topological insulators that are not adiabatically connected to the non-interacting TI limit.
 
In this topical review we will flush out some of the connections between TIs and QSLs recently established.  In this paper, we will focus on a number of different lattice models of fermions.  Some of the models are non-interacting, some are interacting, some have no intrinsic spin-orbit coupling, while some do have intrinsic spin-orbit coupling.  We study a variety of  lattices in two and three spatial dimensions and focus on the interplay of lattice geometry, interactions, and spin-orbit coupling.  Our aim is to enlarge our understanding of the phase diagram of interacting fermions in two and three dimensions by solving different model problems and using them as points of generalization.  For example, which features appear often and therefore are rather general, and which features appear to be model specific?  This approach is complementary to directly developing an effective theory.  Effective theories are extremely powerful because of their generality and efficient exploitation of the various symmetries present in a problem.  However, it is not always straightforward to ``guess" an effective theory of a new phase of matter so special cases derived from more microscopic lattice models can serve as a useful guide, particularly when strong interactions and/or correlations are present.  That is the point-of-view we take in this article.

A summary of our main results is as follows.  For the case of two-dimensional lattices we show that under rather general conditions there is a connection between TIs and Kitaev-type\cite{Kitaev:ap03,Kitaev:ap06} spin models.  An earlier work by Lee, Zhang, and Xiang\cite{Lee:prl07} connecting the honeycomb Kitaev model and the integer quantum Hall effect was an important inspiration for our studies.  A relationship between the integer quantum Hall effect, TIs, and Kitaev models exists because the latter can be exactly solved by a mapping that reduces the interacting spin degrees of freedom to non-interacting Majorana degrees of freedom.\cite{Kitaev:ap06}  These Majorana fermions hop on the lattice and have a band structure similar to ``normal" fermions (only with a redundancy in the labeling of the states).  Therefore, one may study the topological properties of the energy bands, such as whether they possess a finite Chern number.   Energy bands of finite Chern number will turn out to establish a connection between TIs and Kitaev-type spin liquids (and of course the integer quantum Hall effect).  

Staying with two-dimensions, we will discuss recent results on obtaining a fractional quantum Hall effect in a flat-band lattice model with a partially filled band with a finite Chern number.  This lattice model may prove useful in the studies of ``fractional" topological insulators in two-dimensions with time-reversal symmetry.\cite{Bernevig:prl06,Levin:prl09,Qi11,Lu11}  We also describe a class of models in which spin-orbit coupling can be spontaneously generated at the mean-field level.\cite{Raghu:prl08}  We contrast the relative instability to electron interactions of Dirac-like and quadratic band touching points in the non-interacting band structure.  We find that when quadratic band touching points are present, topological phases often appear as the leading instability with interactions.  Such mean-field studies reduce a many-particle problem to a single-particle problem and serve at least two key roles in the study of topological insulators: (1) They elucidate where topological phases can be expected to appear in the presence of electron interactions and (2) They help guide the class of non-interacting lattice models that contain interesting physics. We finish our discussion of two-dimensional systems with an example of theoretical TI studies guiding the discovery of a QSL with novel properties\cite{Chua:prb11} and comment on the role that disorder can play in stabilizing topological quantum spin liquids\cite{Chua11} and topological insulators.\cite{Jain:prl09,Groth:prl09,Prodan:prb11,Guo:prl10,Prodan:prl10}

In the case of three-dimensions, we focus on systems with intrinsic spin-orbit coupling and intermediate strength interactions.  With an eye towards transition metal oxides with 4$d$ and 5$d$ elements, we use the slave-rotor mean-field theory\cite{florens:prb02,florens:prb04} to investigate the competition between interactions, spin-orbit coupling and lattice distortions on the pyrochlore lattice.  We find that a new phase--the ``weak topological Mott insulator"--appears\cite{Kargarian:prb11} with topologically protected gapless modes that carry heat only along a certain class of bulk defects.\cite{Ran:np09}  
As a function of pressure, a complex set of transitions between conventional and exotic phases is possible in such systems.  We discuss the connection of the slave-rotor mean-field states to quantum spin liquids describe by fermionic excitations and suggest some ways the latter may inform the study of interacting TIs in three dimensions.\cite{Fiete:sci11}  We close with a discussion of a newly emerging direction in transition metal oxides: interaction-driven topological phases at interfaces.\cite{Ruegg11_2,Wang11,Yang11,Xiao11}

Our paper is organized as follows.  In Sec. \ref{sec:prelim} we set the convention for our use  of the term ``topological order" in this paper and provide some important general background, including TI-QSL connections in one-dimension and important results for interacting systems.  In Sec. \ref{sec:2D} we detail our main results for two-dimensions, and in Sec. \ref{sec:3D} we describe our chief three-dimensional results.  Finally, in Sec. \ref{sec:conclusions}  we summarize our main results and give our view of interesting future directions of study.

\section{Preliminaries}
\label{sec:prelim}

\subsection{Topological terminology}

Our main goal in this paper is to provide examples of how TIs and QSLs are related to each other, and how the study of each can help deepen our understanding of the other.  An important ``connection" between these two types of phases is via their topological\cite{Volovik} properties.  In Sec. \ref{sec:2D}, we will be more precise by what we mean by ``connection".  We note here that it {\em does not simply mean adiabatic connection} of the electronic states because TIs and QSLs are distinct phases of matter.  In order to help avoid confusion later, we emphasize that we will use ``topological" in the broadest sense of the word: there is some mathematical invariant characterizing the global properties of the system that can not change unless a bulk gap in the energy spectrum closes.  Those with a background in the fractional quantum Hall effect or exotic quantum spin liquids commonly take  ``topological order"  to be synonymous with ``non-trivial ground state degeneracy" and ``non-trivial ({\em i.e.} fractional) quantum numbers/statistics" of excitations.\cite{Wen}  Since a large part of this article deals with TIs which do not have a non-trivial ground state degeneracy, but nevertheless do have topological properties,  we choose a more encompassing use of the terminology.  

For purposes of this article, we would say that $Z_2$ time-reversal invariant TIs, integer quantum Hall systems, fractional quantum Hall systems, and  QSLs described at low energies by a $Z_2$ gauge theory all possess ``topological order".  We find this use of terminology convenient for highlighting the different roles that topological invariants and conventional order parameters play in describing a state of matter.  This way, the quantities characterizing a phase are divided into ``non-local" (topological) and ``local" (conventional order parameter) types.  (As an aside, we note that such divisions can be subtle and sometimes go against conventional wisdom.  For example, superconductors should be classified as topologically ordered in the sense of Wen,\cite{Wen} as they have a non-trivial ground state degeneracy and possess no {\em gauge invariant local order parameter}.\cite{Hansson:ap04})

\subsection{General remarks on one-dimensional systems}

In this topical review, we are mainly concerned with two and three-dimensional systems.  However, a few remarks about one-dimensional systems are in order.  One-dimensional Fermi systems are special for a number of reasons:  (i) They are strongly correlated at low-energies regardless of the strength of the interactions, and they possess interesting properties such as spin-charge separation, a kind of fractionalization of fermion quantum numbers.\cite{Giamarchi,Gogolin}  (ii) Their theoretical study is amenable to powerful, non-perturbative techniques based on Abelian and non-Abelian bosonization techniques.\cite{Giamarchi,Gogolin}  A well-known gapless phase of one-dimensional fermions conveniently described with these techniques is the Luttinger liquid.\cite{haldane}
In recent years it has become clear that interesting, highly universal ``spin-incoherent" Luttinger liquid regimes also appear.\cite{Fiete:rmp07,Fiete:jpcm09,Feiguin:prl11} 

Focusing our attention on gapped one-dimensional systems, a few results are suggestive of possible connections between TIs and QSLs.   While there are no $Z_2$ topological insulators in one dimension,\cite{Ryu:njp10,Kitaev:periodic,Qi:prb08}
 the more general Altland and Zirnbauer classifications\cite{Altland:prb97,Schnyder:prb08}  of non-interacting topological states has provided us with important results for other topological phases resulting from different sets of symmetries.  Of particular interest is the class BDI (time-reversal symmetry with sublattice symmetry and integer spin), which has a $Z$ classification in one dimension.\cite{Ryu:njp10,Kitaev:periodic}  Fidkowski and Kitaev\cite{Fidkowski:prb10} have recently shown in a one-dimensional model in class BDI that the $Z$ classification breaks down to $Z_8$ in the presence of interactions, and have further generalized these results to other symmetry classes.\cite{Fidkowski:prb11}  In other words, phases that are topologically distinct at the non-interacting level can be adiabatically connected with interactions that are slowly turned on and off in a particular way. This example proves that interactions can lead to important modifications of the non-interacting topological classifications.\cite{Gurarie:prb11}   A similar breakdown may occur in two and three-dimensional systems, although to our knowledge there is not yet a concrete example.
 
\subsection{Entanglement as a tool to study topological order}

Because topological order is a global, non-local property, it is rather difficult to measure in the general situation.  (Most physical responses are described by the {\em local} coupling of an order parameter to an external perturbation, or generalized ``force".\cite{Mahan})  In a few important cases, such as the integer and fractional quantum Hall effects, and the $Z_2$ topological insulators, the topological properties give rise to certain quantized responses.\cite{Wen,Qi:prb08,Tse:prl10,Rosenberg:prb10} These examples have topologically protected gapless boundary modes that dominate the responses.  However, often there is no obvious ``nice" response that can be computed or observed in experiment.  The challenge of understanding the many-body quantum mechanics in these cases has led to important advances in the understanding of quantum entanglement in recent years.\cite{Eisert:rmp10,Amico:rmp08} 

In particular, the entanglement entropy\cite{Vidal:prl03,Kitaev:prl06,Levin:prl06} and the entanglement spectrum\cite{Li:prl09} have emerged as two important measures of the quantum entanglement and the topological properties.  Typically, a reduced density matrix of the ground state is computed by tracing over some degrees of freedom (usually real-space coordinates for part of the system).   The topological entanglement entropy, computed by tracing over the reduced density matrix times the logarithm of the reduced density matrix,  contains important information about the ``quantum dimensions" of the excitations in the system.\cite{Kitaev:prl06,Levin:prl06}  This measure is useful for topological systems with excitations that have ``exotic" quantum numbers, such as the fractional quantum Hall effect.\cite{Nayak:rmp08}  (The non-interacting $Z_2$ topological insulators do not have ``exotic" quantum numbers, so the topological entanglement entropy is not a particularly revealing quantity in this case.  However, in a recently identified strongly correlated QSH* phase not adiabatically connected to the QSH phase, the entanglement entropy can serve as a means to distinguish them.\cite{Ruegg11}) 

On the other hand, the entanglement spectrum (ES)\cite{Li:prl09} has proven to be useful in the study of topological insulators.\cite{Turner:prb10,Turner_ES:prb10,Kargarian:prb10,Hughes:10,Prodan:prb11,Thomale:prl10}  The ES is found by writing the reduced density matrix as an exponential of a fictitious Hamiltonian.  The eigenvalues of this fictitious Hamiltonian constitute the entanglement spectrum.  It is remarkable that the ES, which is determined only from the ground-state wave function, contains important information about the {\em excitations} of the system.\cite{Li:prl09}  In particular, for non-interacting models (such as $Z_2$ topological insulators and superconductors) it has been shown that gapless boundary modes imply a gapless ES.\cite{Fidkowski:prl10,Turner:prb10,Turner_ES:prb10}   Moreover, it has recently been proved by Chandran, Hermanns, Regnault, and Bernevig\cite{Chandran11} that in fractional quantum Hall states the low-energy modes of the entanglement spectrum and edge-state excitation spectrum are in one-to-one correspondence.  Qi, Katsura, and Ludwig have obtained similar results.\cite{Qi_entang} 

One important result to emerge from the study of the ES of topological insulators is that a gapless ES can persist under some conditions where the physical edge spectrum becomes gapped.  For example, applying a magnetic field to an inversion symmetric $Z_2$ topological insulator will gap the surface states but leave the entanglement spectrum gapless.\cite{Kargarian:prb10,Turner_ES:prb10} The same is true if time-reversal symmetry is broken in more subtle ways.\cite{Kargarian:prb10}  Strong interactions that gap the edge modes\cite{Wu:prl06,Xu:prb06} will also leave a degeneracy in the ES.\cite{Fu_1D:prb07}  Thus, the ES ``remembers" the underlying state is topological.  In this case, it turns out that the gapless nature of the ES is protected by an inversion symmetry.\cite{Turner:prb10,Turner_ES:prb10,Kargarian:prb10,Hughes:10}  In the case of three-dimensional insulators with inversion symmetry, a gapless entanglement spectrum implies an ``$\bf E\cdot B$" term in the action {\em even if the boundary contains no gapless modes}.\cite{Turner_ES:prb10}  In this sense, the entanglement spectrum can be used to identify a phase of matter.\cite{Hughes:10}  Recently it has been shown that the breakdown of the $Z$ classification (to $Z_8$) of class BDI in the presence of interactions\cite{Fidkowski:prb10,Fidkowski:prb11} can be understood via the ES.\cite{Turner:prb11}

We close this subsection on entanglement by returning once more to one dimension and QSLs.  For aficianados of one-dimensional  spin chains, Haldane's classification\cite{Haldane:prl83}  of antiferromagnetically coupled Heisenberg spins is celebrated: Half-odd integer multiple spin chains are gapless (Luttinger liquids) while integer spin chains are gapped and possess no long-range order, i.e. they are gapped QSLs.  Technically, the two cases are distinguished by the Berry phase (topological) term in the effective low-energy action.  Remarkably, the gapped integer spin systems contain {\em topological order} which manifests itself as a non-local string order parameter with long range order,\cite{denNijs:prb89} and {\em protected gapless boundary modes}\cite{Kennedy:prb92} (similar to TIs).   In fact, ES studies of integer spin chains show phenomenology quite similar to TIs.  For example, the physical edge state degeneracies lead to the same degeneracies in the ES.\cite{Pollmann:prb10}  Additionally, perturbations that destroy long-range order in the string parameter and edge-state degeneracy leave the ES degeneracy in tact, so long as inversion symmetry is preserved.\cite{Pollmann:prb10} The ES results show that the perturbations to spin chains that destroy the conventional measures of the topological order in the Haldane phase, long-range string order and edge-state degeneracies, do not destroy the Haldane phase itself (so long as the bulk gap remains open). (Even in gapless one-dimensional systems the ES may provide useful information.\cite{Thomale_spin:prl10})  

Thus, the entanglement properties have emerged as a robust method of identifying a quantum phase.\cite{Chen:prb11}  It is remarkable that the behaviors of the ES spectrum for TIs and QSLs exhibit such similar phenomenology, both with and without perturbations that destroy the edge modes.  As we will see in the next section, there are more direct ways that TIs and QSLs can be related in higher spatial dimensions.

\section{Two-dimensional systems}
\label{sec:2D}

In this section we will describe a class of non-interacting and interacting lattice models with topological order in two spatial dimensions. Some of these are spin models of QSLs and some are tight-binding models of TIs.  We will show that under a certain set of conditions there is a precise topological connection between them.

\subsection{Orientation}

The first prediction of a quantum spin Hall state (a two-dimensional TI) in a specific material was made by Bernevig, Hughes, and Zhang\cite{Bernevig:sci06} for HgTe quantum wells.  This prediction was soon verified experimentally in a series of beautiful experiments in W\"urzburg.\cite{Konig:sci07,Roth:sci09} 

In this topical review, we will focus on models inspired by the earlier pioneering work of Kane and Mele,\cite{Kane:prl05,Kane_2:prl05} which first established the quantum spin Hall state as a new topological phase in a simple tight-binding model on the honeycomb lattice that preserves time-reversal symmetry.   Their Hamiltonian is
\begin{eqnarray}
\label{eq:K-M}
{\cal H}_{\rm K-M}=-t\sum_{\langle ij\rangle, \sigma}(c_{i\sigma}^\dagger c_{j\sigma}&+& {\rm H.c.})+{\cal H}_{\rm CDW}+{\cal H}_{\rm R}\nonumber \\ &+&i\lambda_{\rm SO}\!\!\!\!\!\!\!\!\sum_{\langle \langle ij\rangle \rangle, \alpha, \beta}\!\!\!\!\!\!{\vec e_{ij}}\cdot {\vec s_{\alpha \beta}} c_{i\alpha}^\dagger  c_{j\beta},
\end{eqnarray}
where $t$ is a real first-neighbor hopping, $c_{i\sigma}^\dagger$ creates a particle of spin $\sigma$ on site $i$ and $c_{i\sigma}$ annihilates the same particle, $\lambda_{\rm SO}$ is the (real) parameter representing the strength of the second-neighbor spin-orbit coupling, ${\vec e_{ij}}=({\bf d}^1_{ij}\times {\bf d}^2_{ij})/|{\bf d}^1_{ij}\times {\bf d}^2_{ij}|$ is a vector normal to the $x-y$ plane describing how the path $\langle \langle i j\rangle \rangle$ was traversed, that is, either ``bending" to the right or left.\cite{Kane:prl05,Kane_2:prl05} Here, ${\cal H}_{\rm CDW}$ and ${\cal H}_{\rm R}$ are charge density wave and Rashba spin-orbit terms, respectively.  The CDW term describes a mean-field staggered sublattice potential that would gap the Dirac points if $\lambda_{\rm SO}=0$ and put the model in a trivial insulator phase.  The Rashba spin-orbit term breaks $S^z$ conservation and is generally present if the physical system has broken inversion symmetry with respect to the plane of the honeycomb lattice.

A key precedent to Kane and Mele's work was a result from Haldane\cite{Haldane:prl88} in which it was shown that a tight-binding model of {\em spinless} fermions on the honeycomb lattice with broken time-reversal symmetry {\em but zero net magnetic field} could exhibit a quantum Hall effect. (In other words, it is the breaking of time-reversal symmetry that is key for the quantum Hall effect, not the net magnetic flux.)  Haldane's model is
\begin{eqnarray}
\label{eq:H}
{\cal H}_{\rm H}=-t\sum_{\langle ij\rangle}(c_{i}^\dagger c_{j}&+& {\rm H.c.})+{\cal H}_{\rm CDW}\nonumber \\ &+&it_2\sum_{\langle \langle ij\rangle\rangle}{\rm sgn}({\vec e_{ij}})c_{i}^\dagger  c_{j},
\end{eqnarray}
where $t_2$ is the (real) parameter describing the string of the second neighbor hopping and sgn(${\vec e_{ij}}$) is $\pm 1$ depending on whether the spinless fermion made a right or left ``turn" along the path $\langle \langle ij\rangle\rangle$.

As Kane and Mele emphasized in their work,\cite{Kane:prl05,Kane_2:prl05}  in a quantum spin Hall system that preserves $S^z$ quantization [${\cal H}_{\rm R}=0$ in \eqref{eq:K-M}], the state can be represented as ``two copies" of Haldane's model \eqref{eq:H}--one for spin up and one for spin down--each of which sees an opposite sign of the effective magnetic field and therefore preserves time-reversal symmetry overall.  The presence of a small but finite Rashba coupling does not change the phase; it only destroys the $S^z$ conservation.

While the topological classification of the quantum Hall effect is based on the Chern number\cite{Nayak:rmp08} (which takes the integers $Z$) and the two-dimensional time-reversal invariant TIs are classified by $Z_2$,\cite{Qi:pt10,Moore:nat10,Hasan:rmp10} the Haldane-Kane-Mele correspondence shows that the integer quantum Hall effect in a lattice model is intimately related to the quantum spin Hall state.  In particular, if a spin polarized tight-binding model realizes a quantum Hall effect, combining two time-reversed copies of it with spin degrees of freedom is guaranteed to produce a quantum spin Hall effect.  This simple observation is actually a powerful tool for {\em generating} desired topological states in two-dimensional lattice models.  We will use this approach in several different ways below, including generating TIs from QSLs,\cite{Ruegg:prb10,Kargarian:prb10} generating QSLs from TIs\cite{Chua:prb11} (the Haldane-Kane-Mele correspondence in reverse), and generating {\em fractional} TIs from the fractional quantum Hall effect in lattice models.\cite{Hu:prb11}

\subsection{``Non-interacting" lattice models}
\label{sec:non-interacting}

After the work of Kane and Mele on the honeycomb lattice\cite{Kane:prl05,Kane_2:prl05} there was a flurry of activity exploring simple non-interacting tight-binding lattice models on a variety of different lattices in two and three spatial dimensions.  For example, in two dimensions the decorated honeycomb lattice,\cite{Ruegg:prb10} the checkerboard lattice,\cite{Sun:prl09} the square-octagon lattice,\cite{Kargarian:prb10} the kagome lattice,\cite{Guo:prb09} the ruby lattice,\cite{Hu:prb11} and others\cite{Weeks:prb10} have been shown to support a TI phase in simple $s$-band models.  If multiple orbitals on each site are used, even the simple square lattice can support a TI.\cite{Bernevig:sci06}  

A natural question to ask is ``What properties of a lattice model would allow it to support a TI phase?"  One way to view this question in light of the discussion above is ``What properties of a spinless fermion lattice model would allow it to support an integer quantum Hall effect?"  When cast this way, there are immediate connections to so-called ``chiral" QSLs, a class of spin liquids with broken time-reversal symmetry and a thermal Hall conductance.  In fact, one of the early models relating spin liquids to the (fractional) quantum Hall effect was proposed by Laughlin himself.\cite{kalmeyer1987}  

In recent years, an important class of exactly solvable spin models known as ``Kitaev" models has emerged.\cite{Kitaev:ap03,Kitaev:ap06}  These models can be used to generate concrete examples of various types of QSLs (previous studies often relied on uncontrolled approximations).  In particular, chiral spin liquids with a quantized thermal Hall conductance can be found.\cite{Yao:prl07,Chua:prb11,Kells:prb10}  As we now show, such models can be adiabatically deformed to an integer quantum Hall ``Haldane-type" model, which in turn can be used to construct a two-dimensional TI via the Haldane-Kane-Mele correspondence described above.

As a simple example, consider the Kitaev model on the honeycomb lattice,\cite{Kitaev:ap06}
\begin{equation}
\label{eq:Kitaev}
{\cal H}_{\rm Kitaev}=\sum_{\langle ij\rangle _x} J \sigma_i^x\sigma_j^x+\sum_{\langle ij\rangle _y} J \sigma_i^y\sigma_j^y+\sum_{\langle ij\rangle _z} J \sigma_i^z\sigma_j^z,
\end{equation}
where the spin-1/2 components $\sigma^x,\sigma^y,\sigma^z$ couple to each other along the $x,y,z$-labeled links, respectively. The symmetry of the Hamiltonian \eqref{eq:Kitaev} is unusual (not Heisenberg, XY, or Ising, for example), but there are recent proposals suggesting it would arise naturally in some solid state contexts.\cite{Wang:prb10,Jackeli:prl09} The main reason for choosing the form \eqref{eq:Kitaev} is that it admits an exact solution with a spin-liquid ground state.\cite{Kitaev:ap06}

The Hamiltonian \eqref{eq:Kitaev} is easily solved by choosing an alternate representation of the spin degrees of freedom:  One can write the spins as 
\begin{equation}
\label{eq:Majorana}
\sigma_j^x=i\eta_j^x\eta_j^0,\;\sigma_j^y=i\eta_j^y\eta_j^0,\;\sigma_j^z=i\eta_j^z\eta_j^0,\;
\end{equation}
where the $\eta_j^0,\eta_j^x,\eta_j^y,\eta_j^z$ are four ``flavors" of Majorana fermions on the site $j$.  They have the properties $\eta^a=(\eta^a)^\dagger$, $(\eta^a)^2=1$ on each site for $a=0,x,y,z$, and they satisfy the anticommutation relations $\{\eta_i^a,\eta_j^b\}=2\delta_{ab}\delta_{ij}$.  They contain ``half" the degrees of freedom of a ``normal" fermion (one that satisfies the usual anticommutation relations). Importantly, the transformation \eqref{eq:Majorana} preserves the spin commutation relations $[\sigma_j^a,\sigma_j^b]=2i \epsilon^{abc}\sigma_j^c$ provided we impose the constraint  $D_j\equiv- \eta_j^0\eta_j^x\eta_j^y\eta_j^z=1$ on each site. (The Majorana representation \eqref{eq:Majorana} actually doubles the Hilbert space, so one needs to project to the physical Hilbert space.)  Under the representation \eqref{eq:Majorana} the Kitaev Hamiltonian is transformed to
\begin{eqnarray}
\label{eq:Kitaev_Maj}
{\cal H}_{\rm Kitaev}=\sum_{\langle ij\rangle_x} iJ u_{ij} \eta^0_i\eta^0_j\hspace{3cm}\nonumber \\
+\sum_{\langle ij\rangle_y} iJ u_{ij} \eta^0_i\eta^0_j+\sum_{\langle ij\rangle_z} iJ u_{ij} \eta^0_i\eta^0_j,
\end{eqnarray}
where $u_{ij}=-i\eta^a_i\eta^a_j$ is a $Z_2$ gauge field with eigenvalues $\pm 1$ defined along an $a$-link.  The product of six different $u_{ij}$ around hexagonal plaquettes on the honeycomb lattice defines a $Z_2$ gauge flux through that plaquette, $W_p=\prod u_{ij}$.  The usefulness of the representation \eqref{eq:Majorana} is that now
\begin{equation}
[u_{ij},u_{kl}]=[u_{ij},\;{\cal H}_{\rm Kitaev}]=[W_p,{\cal H}_{\rm Kitaev}]=0,
\end{equation}
so that the $u_{ij}$ are {\em non-dynamical} gauge fields and $W_p$ are {\em non-dynamical} gauge fluxes.  In other words, the $u_{ij}$ in \eqref{eq:Kitaev_Maj} can be taken to be constants, rather than operators where the value of the constants are determined (up to gauge equivalent choices) by the fluxes $W_p$ around the plaquettes in the lattice.  For a bipartite lattice (like the honeycomb lattice) it is known that the uniform flux configuration is lowest in energy,\cite{Lieb:prl94} and for non-bipartite lattices detailed numerical studies indicate the same.\cite{Chua:prb11,Yao:prl07}

With the $u_{ij}$ as constants, it is clear the interacting spin Hamiltonian \eqref{eq:Kitaev} has now been reduced to a {\em non-interacting hopping Hamiltonian} \eqref{eq:Kitaev_Maj}. (Recall that $\eta^a=(\eta^a)^\dagger$.)  Of course, a non-interacting Hamiltonian can always be solved exactly.
In the present case, it useful to Fourier transform to momentum space as one would for any tight-binding Hamiltonian and determine the energy bands $E(\vec k)$.  Because of the self-adjoint property of the Majorana fermions, half of the states are redundant and $E(\vec k)=-E(-\vec k)$.  One ``unusual" property of the tight-binding model \eqref{eq:Kitaev_Maj} is that the hopping amplitudes, $iJu_{ij}$, are purely imaginary (assuming $J$ is real).  

The procedure for solving any Kiteav-type model on any lattice is exactly the same as above, even if the lattice is different and/or the spin is larger.\cite{Chua:prb11,Yao:prl09,Wu:prb09,Chern:prb10,Ryu:prb09,Mandal:prb09}  Staying with the honeycomb lattice model for the moment, it is worth emphasizing that Kitaev's solution also determines whether the ground state of the spin system spontaneously breaks time-reversal symmetry.\cite{Kitaev:ap06}  The statement is the following:  If the lattice contains plaquettes with only an even number of sides (such as the honeycomb lattice) the ground state will not break time-reversal symmetry.  If there are any plaquettes with an odd number of links (such as the decorated honeycomb lattice\cite{Yao:prl07} or kagome lattice\cite{Chua:prb11}) the ground state will spontaneously break time-reversal symmetry.\cite{Kitaev:ap06} This statement follows because time-reversal symmetry sends $u_{ij}\to -u_{ij}$, which will change the sign of $W_p$ on odd-sided plaquettes, but not even-sided ones.

Applying this to the honeycomb lattice model \eqref{eq:Kitaev_Maj} we can immediately infer that since the model does not break time-reversal symmetry, the band structure must have zero Chern number.  On the other hand, the addition of the term\cite{Lee:prl07}
\begin{equation}
\label{eq:3spin}
{\cal H}'=\sum_{(ijk) \in \triangle}J' \sigma_i^y\sigma_j^z\sigma_k^x+\sum_{(ijk) \in \nabla}J' \sigma_i^x\sigma_j^z\sigma_k^y,
\end{equation}
explicitly breaks time-reversal symmetry and leads to a {\em second-neighbor} hopping term
\begin{equation}
\label{eq:3spin_Maj}
{\cal H}'=\!\!\!\!\!\!\sum_{\langle\langle ik\rangle\rangle\in \triangle}\!\!\!\!\!\!-iJ'u_{ij}u_{jk} \eta_i^0\eta_k^0+\!\!\!\!\!\!\sum_{\langle\langle ik\rangle\rangle\in \nabla}\!\!\!\!\!\!iJ'u_{ij}u_{jk} \eta_i^0\eta_k^0,
\end{equation}
where $\triangle$ and $\nabla$ denote up and down pointing triangles of $(ijk)$, with $i$ always taken to be the left most point.  Remarkably, if the Majorana band structure of the Hamiltonian \eqref{eq:Kitaev_Maj} with \eqref{eq:3spin_Maj} are taken to represent spinless fermions (as opposed to Majorana fermions), the states can be adiabatically deformed into the Haldane model \eqref{eq:H} (which contains purely real first-neighbor hopping and imaginary second-neighbor hopping) on the honeycomb lattice.\cite{Lee:prl07}  This implies that the Hamiltonian \eqref{eq:Kitaev_Maj} with \eqref{eq:3spin_Maj} is a topological chiral spin liquid with Chern number $\pm 1$!  

This result establishes a Kitaev-Haldane-Kane-Mele correspondence on the honeycomb lattice that generalizes the earlier Haldane-Kane-Mele correspondence to include chiral spin liquids with finite Chern numbers (provided the chiral spin liquids can be described by fermions with a band structure--there is not such an obvious connected to spin liquids described by a bosonic representation of the spin degrees of freedom).  As we now show, the presence of a chiral spin liquid with finite Chern number on the honeycomb lattice\cite{Lee:prl07} when there is also an integer quantum Hall effect\cite{Haldane:prl88} and a TI phase\cite{Kane:prl05,Kane_2:prl05} is a special case of a more general result. 

\begin{figure}[h]
\includegraphics[width=6cm]{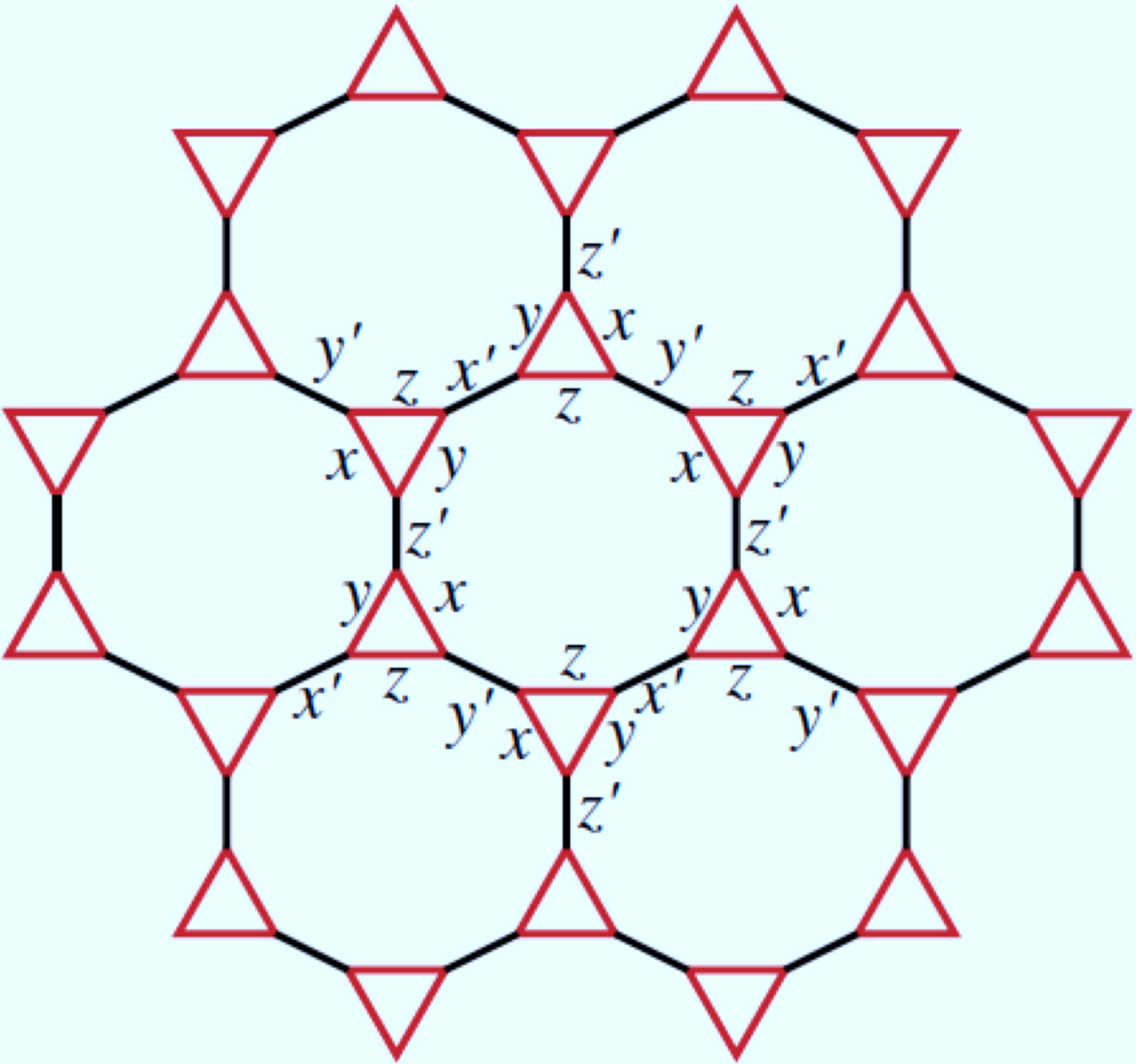}
\caption{(color online) The decorated honeycomb (sometimes called the ``star") lattice.  An exactly solvable chiral spin liquid model with a finite Chern number and non-Abelian anyons has been found on this lattice.\cite{Yao:prl07}  By the Kitaev-Haldane-Kane-Mele correpondence, this implies a topological insulator phase in a simple $s$-band model also exists on this lattice.\cite{Ruegg:prb10}} \label{fig:decorated_Kitaev}
\end{figure}

The first exactly solvable model of a chiral spin liquid with a finite Chern number that included only two-spin interactions was proposed by Yao and Kivelson\cite{Yao:prl07} on the decorated honeycomb lattice shown in Fig.~\ref{fig:decorated_Kitaev}.  The Hamiltonian is of the Kitaev type and is given by
\begin{eqnarray}
\label{eq:decorated_Kitaev}
{\cal H}_{\rm dhc}=\sum_{\langle ij\rangle_x} J \sigma_i^x\sigma_j^x+\sum_{\langle ij\rangle_y} J \sigma_i^y\sigma_j^y+\sum_{\langle ij\rangle_z} J \sigma_i^z\sigma_j^z\nonumber \\
+\sum_{\langle ij\rangle_{x'}} J' \sigma_i^x\sigma_j^x+\sum_{\langle ij\rangle_{y'}} J' \sigma_i^y\sigma_j^y+\sum_{\langle ij\rangle_{z'}} J' \sigma_i^z\sigma_j^z,\,
\end{eqnarray}
where the exchange is given by $J$ along the links of the triangles and $J'$ along the links between triangles, as shown in Fig.~\ref{fig:decorated_Kitaev}.  The Hamiltonian \eqref{eq:decorated_Kitaev} is solved using the transformation \eqref{eq:Majorana}  and a fermion hopping model with purely imaginary hopping parameters results, similar to what is found in \eqref{eq:Kitaev_Maj}.  Because of the triangular plaquettes on the lattice, the ground state spontaneously breaks time-reversal symmetry.  For $0< J'/J<\sqrt{3}$, the ground state is a gapped chiral spin liquid with Chern number $\pm 1$, while for $J'/J>\sqrt{3}$, the ground state is gapped chiral spin liquid with zero Chern number.\cite{Yao:prl07}  If the Kitaev-Haldane-Kane-Mele correspondence described above on the honeycomb lattice is to prove general, Yao and Kivelson's discovery of the chiral spin liquid with finite Chern number in a Kitaev model should imply there is a TI in an $s$-band model on the decorated honeycomb lattice.  We have shown this is indeed the case.\cite{Ruegg:prb10}

\begin{figure}[th]
\includegraphics[width=6cm]{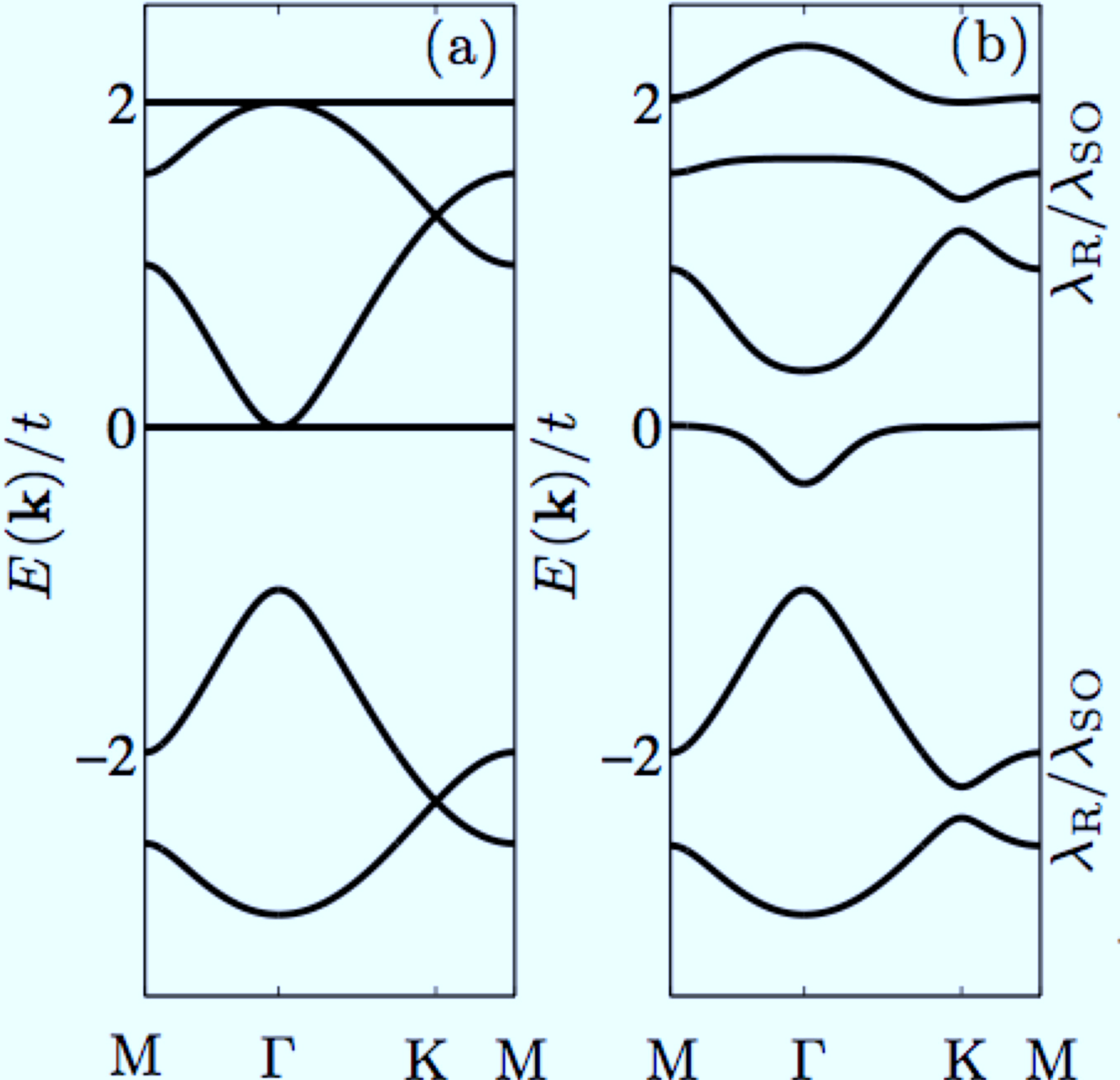}
\caption{The decorated honeycomb lattice energy bands along high symmetry directions in the Brillouin zone (from Ref.[\onlinecite{Ruegg:prb10}]). In (a) $\lambda_{\rm SO}=0$ and in (b) $\lambda_{\rm SO}=0.1t$.  There are Dirac points at $K$ and $K'$ (not shown) and quadratic band crossing points (QBCP) at $\Gamma$ in (a), while in (b) $\lambda_{\rm SO}\neq 0$ opens up a gap at each of these points and destroys the flat bands. } \label{fig:dhc_bands}
\end{figure}

\begin{figure}[th]
\includegraphics[width=8cm]{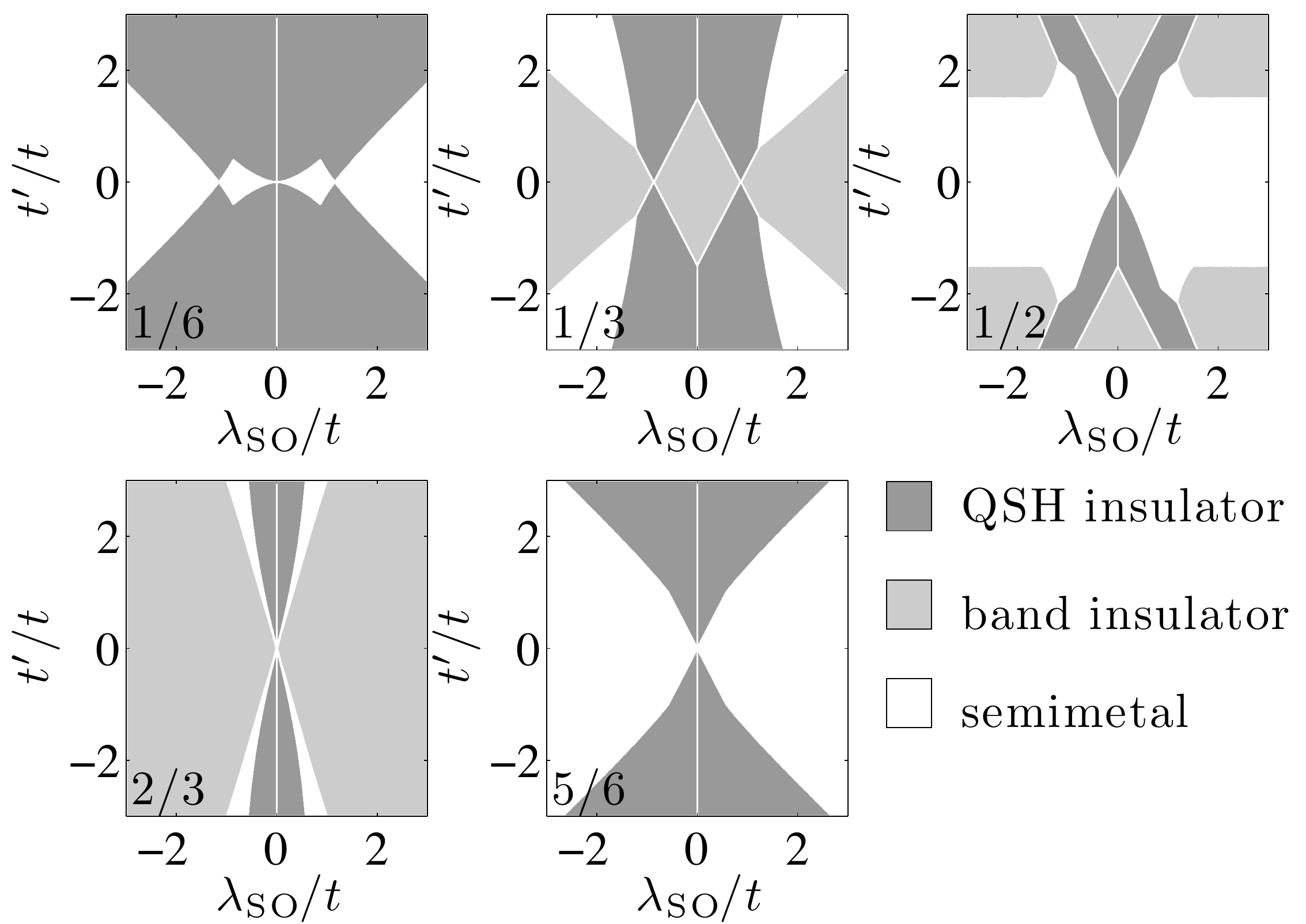}
\caption{Phase diagrams for the decorated honeycomb lattice with $t$ and $t'$ real in the absence of a staggered on-site potential and no Rashba coupling (from Ref.[\onlinecite{Ruegg:prb10}]). Several filling fractions $f$ are shown (lower left corner). For fixed $f$ and $\lambda_{\rm SO}$ it is possible to  drive a transition between a topological insulator and a non-topological phase by varying the ratio $t'/t$.} \label{fig:dhc_phases}
\end{figure}

Consider the $s$-band Hamiltonian on the decorated honeycomb lattice,
\begin{eqnarray}
\label{eq:dhc-TB}
{\cal H}_{\rm dhc-TB}=-t\!\! \sum_{\langle i j\rangle,\sigma,\Delta} c^\dagger_{i\sigma}c_{j\sigma}
-t' \!\!\!\!\!  \sum_{\langle i j\rangle,\sigma,\Delta \to \Delta} \!\!\!\!\! c^\dagger_{i\sigma}c_{j\sigma}+{\rm H.c.},\nonumber \\
+{\cal H}_{\rm CDW}+{\cal H}_{\rm R}+i\lambda_{\rm SO}\!\!\! \sum_{\langle \langle i j\rangle \rangle, \alpha, \beta} \!\!\! \vec e_{ij}\cdot {\vec s}_{\alpha \beta}  c^\dagger_{i\alpha}c_{j\beta},\hspace{.5cm}
\end{eqnarray}
which describes nearest-neighbor hopping on the triangles ``$\Delta$" with amplitude $t$ and between triangles ``$\Delta \to\Delta$" with amplitude $t'$, and the remaining terms describe the same physics as represented in \eqref{eq:K-M}. The bands for ${\cal H}_{\rm CDW}={\cal H}_{\rm R}=0$ are shown in Fig.~\ref{fig:dhc_bands}, and the phase diagrams at various filling fractions  are shown in Fig.~\ref{fig:dhc_phases}.  Note that at all filling fractions there is a finite region of parameter space occupied by a TI (QSH) phase. 

\begin{figure}[th]
\includegraphics[width=8cm]{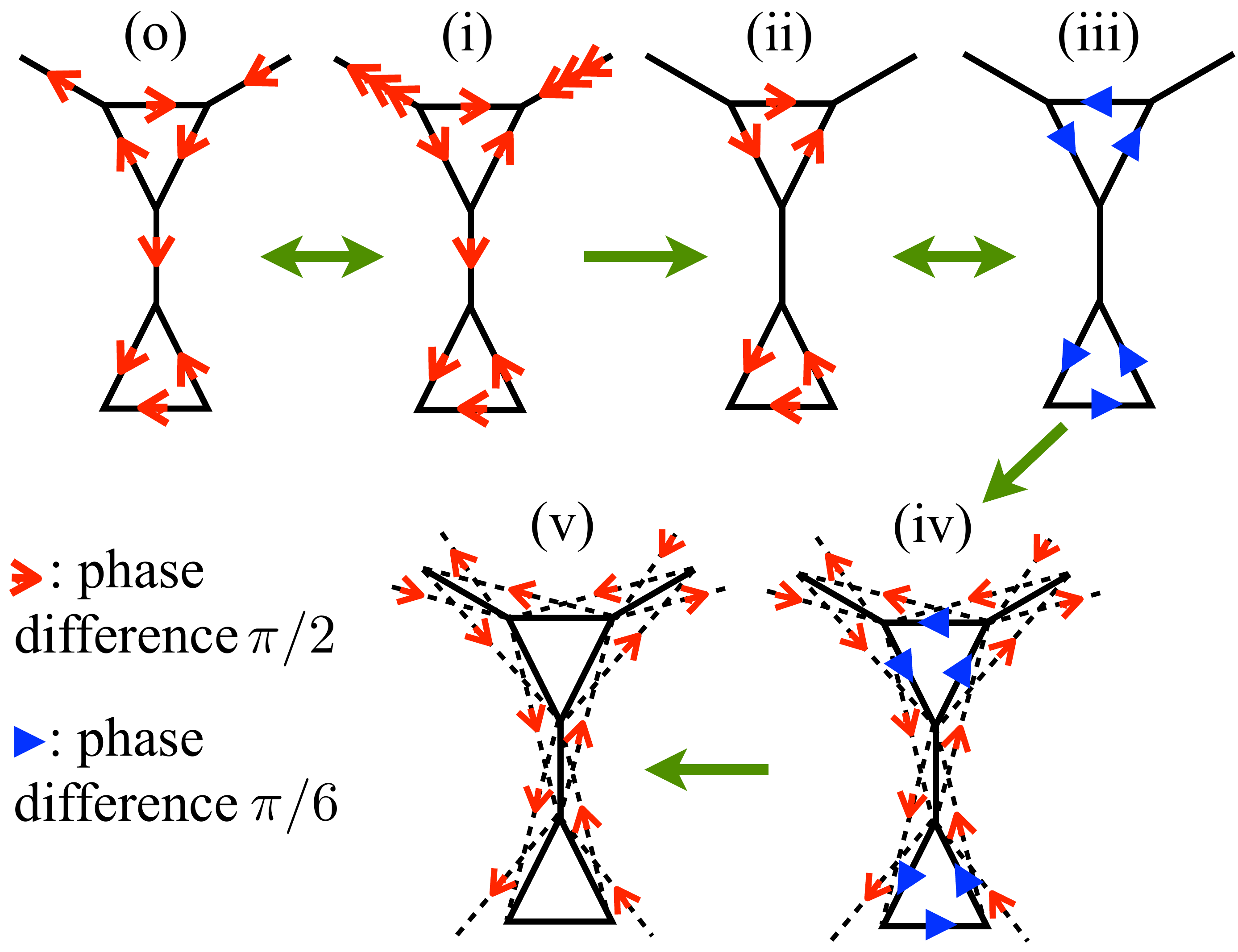}
\caption{(color online) (From Ref.[\onlinecite{Ruegg:prb10}].) Schematic illustration of the continuous path which adiabatically connects the representative free fermion model of the Kitaev model on the decorated Honeycomb lattice with the spinless model with real $t$, $t'$ and $\lambda_{\rm SO}$. The patterns (o) and (i) are equivalent and correspond to Kitaev's model whereas (v) represents the model with real $t$, $t'$ and $\lambda_{\rm SO}$. The adiabatic deformation does not lead to a gap closing and the Chern number stays constant.  See Fig.~\ref{fig:ChernAndGap}. This establishes the topological connection between the two models.} \label{fig:amp_tun}
\end{figure}

\begin{figure}[h]
\includegraphics[width=8cm]{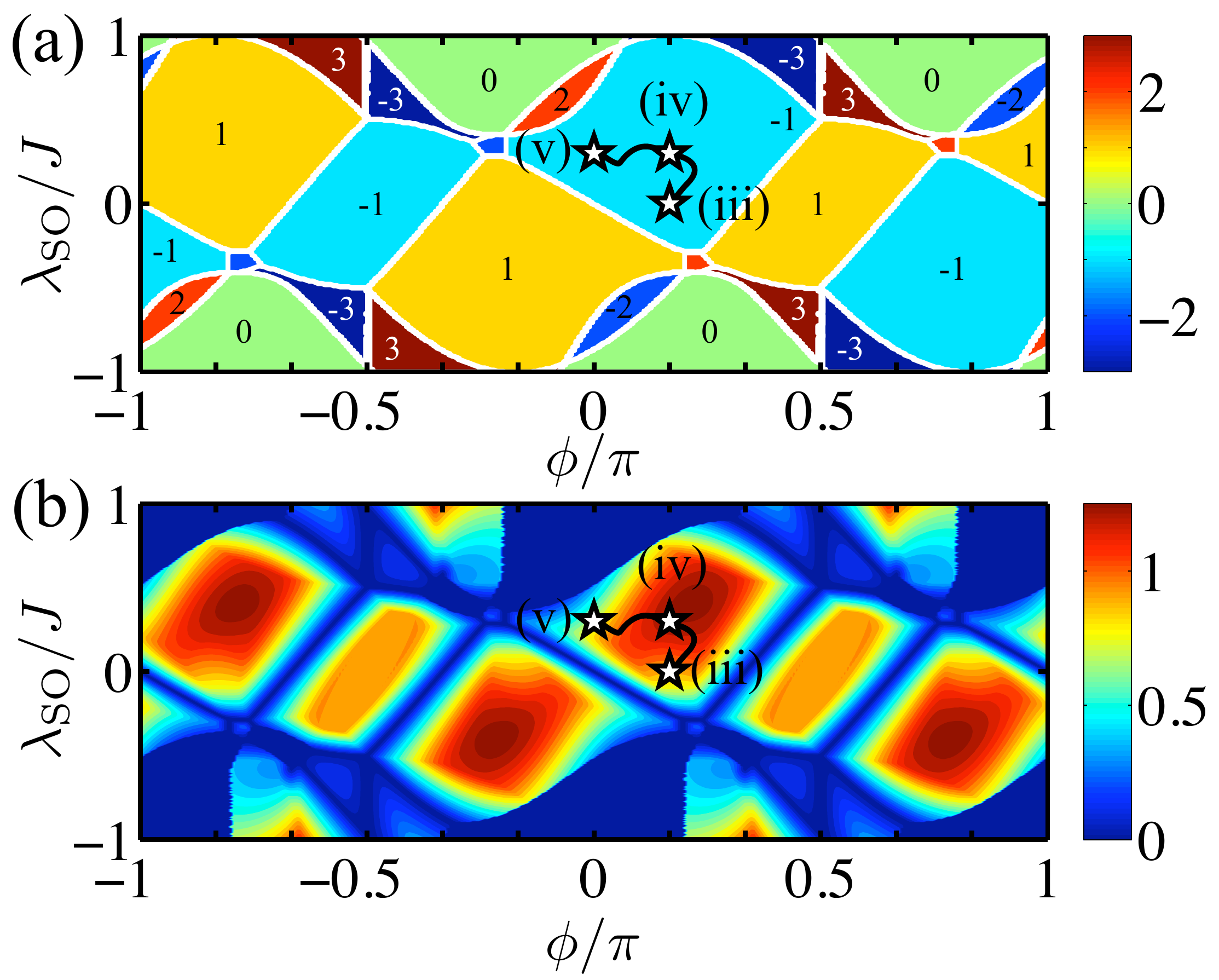}
\caption{(color online) Contour plot of (a) the Chern number and (b) the gap of the model as a function of ($\phi,\lambda_{\rm SO}$) at half-filling for $J=J'$.  Here $\phi$ is the phase angle on the nearest neighbor hopping parameters given in Ref.[\onlinecite{Ruegg:prb10}].  Also shown is a possible path which adiabatically connects the flux patterns (iii), (iv), and (v) defined in Fig.\ref{fig:amp_tun}.} \label{fig:ChernAndGap}
\end{figure}

We now show that the TI phase at 1/2 filling is ``connected" to the chiral spin liquid phase of \eqref{eq:decorated_Kitaev} via the Kitaev-Haldane-Kane-Mele correspondence demonstrated on the honeycomb lattice by Lee, Zhang, and Xiang.\cite{Lee:prl07}  We do this by showing that the Hamiltonian describing the band structure of \eqref{eq:decorated_Kitaev} (with ``normal" spinless fermions replacing the Majorana fermions) with only purely imaginary first-neighbor hopping can be adiabatically deformed to one spin component of  \eqref{eq:dhc-TB} (with ${\cal H}_{\rm CDW}={\cal H}_{\rm R}=0$) which has purely real first-neighbor hopping and purely imaginary second-neighbor hopping.  This is done by applying a combination of gauge transformations and adiabatic tuning of the real and imaginary components of first and second neighbor hopping parameters.  The procedure is shown in Fig.~\ref{fig:amp_tun}, and the panels in Fig.~\ref{fig:ChernAndGap} explicitly show that the Chern number remains unchanged and the gap remains open in this process, establishing the topological equivalence, and the Kitaev-Haldane-Kane-Mele correspondence on the decorated honeycomb lattice.\cite{Ruegg:prb10}

It should be clear from the examples of the decorated honeycomb lattice and the honeycomb lattice above that the Kitaev-Haldane-Kane-Mele correspondence is rather general. However, it is also true that one can invert the argument to use the existence of a TI in an $s$-band model on a particular lattice to infer a chiral spin liquid with a finite Chern number in a Kitaev-type model. Indeed, we have used the work of Guo and Franz that established a TI on the kagome lattice\cite{Guo:prb09} to motivate the study of a Kitaev-type model on the same lattice.\cite{Chua:prb11} Not only did we find a new exactly solvable chiral spin liquid model with finite Chern number, but also a phase that possess a stable spin Fermi surface.\cite{Chua:prb11} We note that unstable spin Fermi surfaces have been reported in other models.\cite{Yao:prl09,Baskaran:09,Tikhonov:prl10}

Because the kagome lattice has sites with four links (as opposed to three on the honeycomb, decorated honeycomb, and decorated square lattice), the effective spin in a Kitaev-type model must be at least 3/2.\cite{Wu:prb09}  These higher-spin Kitaev models require that the spin-1/2 Pauli spin matricies be generalized to Gamma matricies.\cite{Yao:prl09}  The spin Hamiltonian we study\cite{Chua:prb11} on the kagome lattice is 
\begin{eqnarray}
\label{eq:H_gen}
{\cal H}  &=&  \Ju\sum_{\ij\in\triangle}\Gamma_{i}^{1}\Gamma_{j}^{2}+\Jd\sum_{\ij\in\nabla}\Gamma_{i}^{3}\Gamma_{j}^{4}\nonumber +J_{5}\sum_{i}\Gamma_{i}^{5}\\
&&+\Ju'\sum_{\ij\in\triangle}\Gamma_{i}^{15} \Gamma_{j}^{25} +\Jd'\sum_{\ij\in\nabla}\Gamma_{i}^{35}\Gamma_{j}^{45},
 \end{eqnarray}
where we have distinguished the nearest neighbor couplings $J_{ij}$ as $\Ju$,
$\Jd$ and $J'_{ij}$ as $\Ju'$ and $\Jd'$ based on whether the link $\ij$ belongs to an up ($\triangle$) triangle or down ($\nabla$) triangle (the kagome lattice can be viewed as a corner sharing network of up and down triangles), and $\ij$ is taken in the counterclockwise
sense for each triangle.  Locally, the five $\Gamma$-matricies satisfy a Clifford algebra, $\{\Gamma_i^a,\Gamma_i^b\}=2\delta^{ab}$, where $a,b=1,...,5$, 
and $\Gamma^{ab}\equiv [\Gamma^a,\Gamma^b]/(2i)$.  In terms of the components of the spin $S=3/2$ operators, \cite{Murakami:prb04,Yao:prl09}
\begin{eqnarray}
\label{eq:Gamma_S}
\Gamma^1=\frac{1}{\sqrt 3}\{S^y,S^z\}, \Gamma^2=\frac{1}{\sqrt 3}\{S^z,S^x\}, 
\Gamma^3=\frac{1}{\sqrt 3}\{S^x,S^y\},\nonumber\\
\Gamma^4=\frac{1}{\sqrt 3}[(S^x)^2-(S^y)^2],~ 
\Gamma^5=(S^z)^2-\frac{5}{4}.\qquad~~
\end{eqnarray}
With the identification \eqref{eq:Gamma_S}, it is clear the model \eqref{eq:H_gen} has a global Ising spin symmetry under 180$^\circ$ rotations about the $z$-axis, and possesses time-reversal symmetry (TRS) (although TRS will be spontaneously broken in the ground state from the triangular plaquettes as we described earlier) in addition to the translational and threefold rotational lattice symmetry mentioned above.

\begin{figure}[h]
\includegraphics[width=7cm]{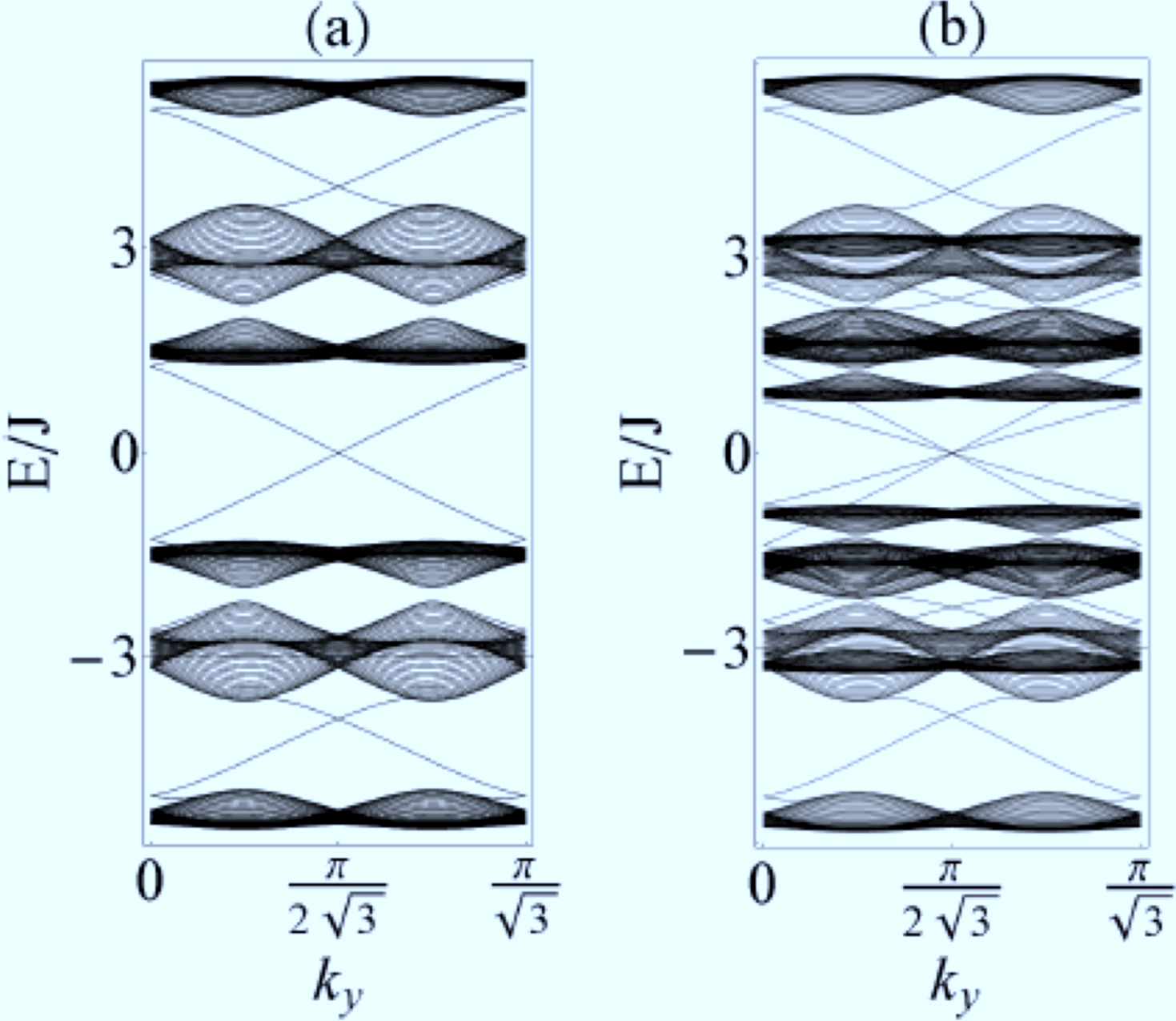}
\caption{(a) Band structure on a cylindrical geometry for $\Ju=\Ju'=1.0$, $\Jd=\Jd'=0.8$, $J_5=0$. There are two gapless chiral Majorana edge states (dotted lines) which overlap on each other. (b) For $\Ju=1.0$, $\Ju'=0.6$, $\Jd=0.9$, $\Jd'=0.5$, and $J_5=0.1$, the two gapless edges states separate. These ground states are thus CSLs with a spectrum Chern number ($\pm 2$) and the vortices obey Abelian statistics.  From Ref.[\onlinecite{Chua:prb11}].} \label{fig:kagome_strip}
\end{figure}

\begin{figure}[h]
\includegraphics[width=5cm]{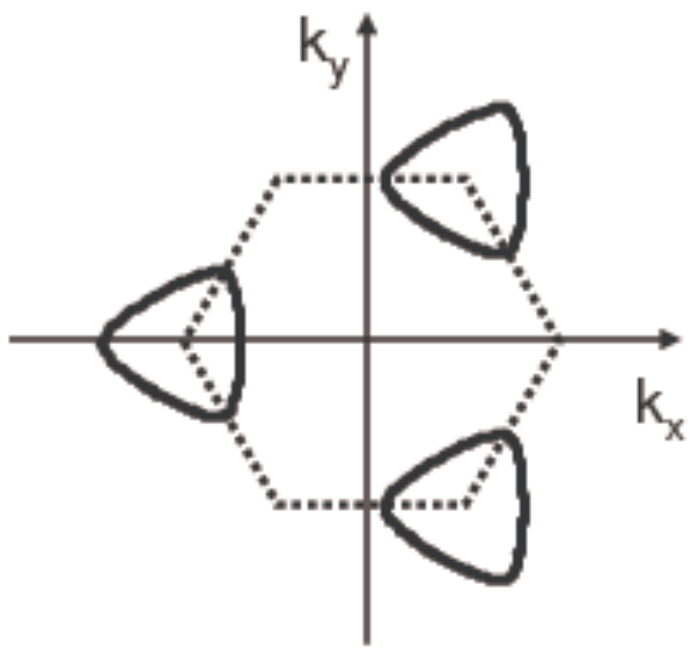}
\caption{The Fermi surface (solid line) for the flux configuration $\{\frac{\pi}{2},\frac{\pi}{2},\pi\}$ and  $\{\Ju,\Jd,\Ju',\Jd',J_5\}=\{1.0, 0.3, 0.8, 0.5,1.4\}$. (See Ref.[\onlinecite{Chua:prb11}] for a description of notation.) The dashed hexagon is the Brillouin zone boundary.  Note that there is only one Fermi pocket for this set of parameters and the three pockets shown are related by a reciprocal lattice vector and are thus equivalent. } 
\label{fig:fermi2}
\end{figure}

In order to solve the higher-spin $\Gamma$-matrix model \eqref{eq:H_gen}, a Majorana representation similar to that for spin-1/2, \eqref{eq:Majorana}, is employed 
 \begin{equation}
 \label{eq:Gamma_Majorana}
\Gamma_{i}^{a}  =  i\xi_{i}^{a}c_{i},\quad
\Gamma_{i}^{5}  =  ic_{i}d_{i},\quad
\Gamma_{i}^{a5}  =  i\xi_{i}^{a}d_{i},
\end{equation}
with $a=1,2,3,4.$  There are thus 6 Majorana species on each site $i$: $\{\xi_{i}^{1},\xi_{i}^{2},\xi_{i}^{3},\xi_{i}^{4},c_{i},d_{i}\}$.  The Majorana representation enlarges the spin-3/2 Hilbert space, so that one must enforce the constraint $D_{i}=-\Gamma_i^1 \Gamma_i^2\Gamma_i^3\Gamma_i^4\Gamma_i^5 =-i\xi_{i}^{1}\xi_{i}^{2}\xi_{i}^{3} \xi_{i}^{4}c_{i}d_{i}=1$, namely for any physical state $|\Psi\rangle_\textrm{phys}$ $D_i|\Psi\rangle_\textrm{phys}= |\Psi\rangle_\textrm{phys}$ for any $i$. The remainder of the steps are similar to the spin-1/2 version and one obtains a Majorana tight-binding model on the kagome with imaginary hopping parameters.\cite{Chua:prb11}  Solving the model on a strip geometry in the gapped phase with a finite Chern number ($\pm 2$) yields the results shown in Fig. ~\ref{fig:kagome_strip}.  The {\em gauge dependent} but {\em stable} Fermi surface in a gapless phase is shown in Fig.~\ref{fig:fermi2}. 

The results of this subsection have demonstrated the Kitaev-Haldane-Kane-Mele correspondence and illustrated how TIs can be generated from a certain class of QSLs, and vice-a-versa. Thus, we have established a {\em method} of generating certain types of desired states.  Moreover, along the way we have also discovered new QSLs, such as the gapless varieties with a stable spin fermi surface.\cite{Chua:prb11}  In all these ``non-interacting" models,  the topological relationship between the TI and QSL models is based on the Chern number and an underlying band structure of free fermions.  In the next subsection we turn to the study of interacting models in two dimensions. 

As a final remark, we note there are also deep connections between Kitaev models, the quantum Hall effect, topological insulators, and topological superconductivity.\cite{DeGottardi:njp11,Grover:prl08,Qi:prl09,Sato:prb09,Fidkowski:prl10,Nayak:rmp08,Schnyder:prl09,Kitaev:ap06}  The topological Kitaev models\cite{Chua11} and the TI insulators (including the integer quantum Hall effect)\cite{Jain:prl09,Groth:prl09,Prodan:prb11,Guo:prl10,Prodan:prl10} show that disorder can play a useful role in {\em stabilizing} topologically ordered states and could play a role in guiding the experimental discovery of topological quantum spin liquids.\cite{Chua11} The non-interacting models discussed in this section can be viewed as mean-field states in a more general interacting model.

\subsection{Interacting lattice models}

While the Kitaev models are interacting in terms of the underlying electronic degrees of freedom (local moments), we discussed them under the heading of ``non-interacting" lattice models because they admit a free (Majorana) fermion representation and therefore can be understood in terms of the properties of their band structure. There are few exactly solvable interacting models in two dimensions, and those that exist often possess special symmetries.\cite{Huijse:prb11}  Thus, one must look for approximate methods to address the physics of interactions.   

For a certain class of Hamiltonians (those without a fermion sign problem), quantum Monte Carlo is a powerful theoretical method.  Given the central role the honeycomb lattice has played in the understanding of TIs and QSLs, it is natural to begin an investigation of interacting models on this lattice.  In the last year the simplest interacting model, the Hubbard model (no spin-orbit coupling terms), has been shown to possess a (apparently gapped) spin-liquid phase at half-filling and intermediate coupling.\cite{Meng:nat10} Moreover, this spin liquid has been shown to be robust in the presence of a second-neighbor spin-orbit coupling in quantum Monte Carlo\cite{Hohenadler:prl11} and dynamical mean-field\cite{Wu:prl11} studies of the so-called Kane-Mele-Hubbard model.   Related quantum Monte Carlo studies on bulk and edge instabilities in the Kane-Mele-Hubbard model have been reported,\cite{Wu:prl11,Zheng:arxiv10,Yamaji:prb11,Yu:prl11} as well as closely related exact diagonalization studies in spinless systems with nearest-neighbor interactions for small system sizes.\cite{Varney:prb10}

Aside from bosonization studies of interactions in the one-dimensional helical liquids that exist along the boundary of a two-dimensional TI,\cite{Strom:prl09,Huo:prl09,Xu:prb10,Teo:prb09,Law:prb10,Zyuzin:prb10,Tanaka:prl09,Maciejko:prl09} most studies of  interactions in bulk have relied on some type of mean-field theory.\cite{Raghu:prl08,Sun:prl09,Wen:prb10,Zhang:prb09,Liu:prb10,Ruegg:prb11,Ruegg11}  One of the most important results to emerge from these studies is that topological insulators can be {\em spontaneously} generated from interactions at the mean-field level {\em even when there is no microscopic spin-orbit coupling present}. This was first explicitly shown by Raghu {\it et al.} on the honeycomb lattice.\cite{Raghu:prl08}  In model without any intrinsic spin-orbit coupling, TI phases are produced by the spontaneous generation of spin-orbit coupling from first and second-neighbor interactions treated at the mean-field level.\cite{Raghu:prl08,Sun:prl09,Wen:prb10,Zhang:prb09,Liu:prb10,Ruegg11_2} 

The stability of a non-interacting band structure to interactions depends to a large extent on the form of the low-energy dispersions and band crossing points.  For example, it is well-known that at 1/2 filling on the honeycomb lattice the non-interacting dispersions are Dirac-like.\cite{Neto:rmp09} Dirac points are perturbatively stable to interactions.\cite{Sun:prl09,Raghu:prl08,Wen:prb10,Weeks:prb10}  In practice, this implies that a critical interaction strength is required to drive a phase transition away from the gapless state.\cite{Sun:prl09}  This is qualitatively consistent with both mean-field studies\cite{Raghu:prl08,Wen:prb10,Rachel:prb10} and quantum Monte Carlo studies.\cite{Meng:nat10} By contrast, two bands that touch quadratically or flatter are {\em infinitesimally} unstable to interactions in two dimensions.\cite{Sun:prl09,Wen:prb10}  Quadratic band touching points are known to occur on many of the lattices that support TIs in simple $s$-band models: decorated honeycomb,\cite{Ruegg:prb10} kagome,\cite{Guo:prb09}  checkerboard,\cite{Sun:prl09} square-octagon,\cite{Kargarian:prb10} and ruby.\cite{Hu:prb11} As we show below, the cases of quadratic band touchings points often have topological phases as the leading instabilities with respect to interactions.\cite{Wen:prb10}  This effectively enlarges the candidate systems for TI physics to include those {\em without strong spin-orbit coupling} but with ``flat" band touching points in the non-interacting band structure.  Another notable example is bilayer and multilayer graphene.\cite{FZhang:prl11} 

The kagome lattice serves as an important example case.  Its non-interacting band structure in an $s$-band nearest neighbor hopping model possess both Dirac points (1/3 filling) and quadratic band touching points (2/3 filling).\cite{Guo:prb09} We study the following Hamiltonian\cite{Wen:prb10}
\begin{eqnarray}
&&{\cal H}_{\rm kagome}=-t\sum\limits_{\langle i,j \rangle}c_{i\sigma}^{\dag }c_{j \sigma}+U\sum\limits_{i}n_{i\uparrow }n_{i\downarrow }\nonumber\\
&&+V_{1}\sum_{\langle i,j \rangle }n_{i }n_{j }
+V_{2}\!\!\sum_{\langle\!\langle i,j \rangle\! \rangle }n_{i }n_{j }+V_{3}\!\!\sum_{\langle\! \langle\!\langle i,j \rangle\! \rangle\! \rangle }n_{i }n_{j }
\label{eq:H_kagome}
\end{eqnarray}
at the mean-field level.  Here, $c_{i\sigma}^{(\dag)}$ annihilates (creates) a fermion on site $i$ with spin $\sigma=\uparrow, \downarrow$, $n_{i\sigma}=c_{i\sigma}^{\dag}c_{i\sigma}^{}$ and $n_i=\sum_{\sigma}n_{i\sigma}$. The sums run over nearest-neighbor $\langle i,j\rangle$, second-neighbor $\langle\!\langle i,j\rangle\!\rangle$, or third-neighbor bonds $\langle\!\langle\!\langle i,j\rangle\!\rangle\!\rangle$. The hopping amplitude is denoted by $t$ and the parameters $V_1$, $V_2$, and $V_3$ quantify the nearest-neighbor, second-neighbor and third-neighbor repulsion, respectively.

\begin{figure}[th]
\centering
\includegraphics[width=0.9\linewidth]{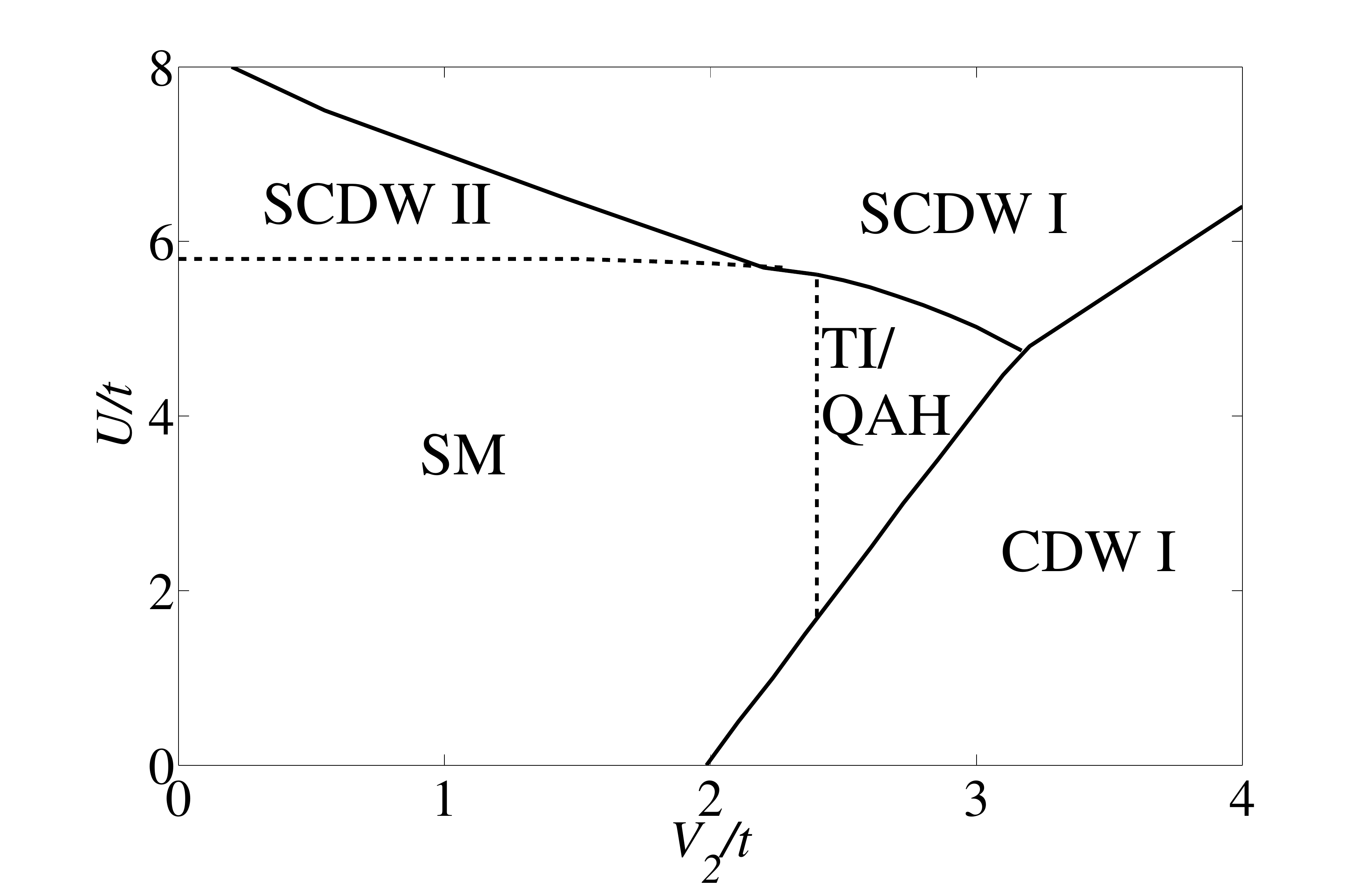}
\caption{Filling fraction 1/3 on the kagome lattice with Dirac points in the non-interacting band structure. The $U$-$V_2$ phase diagram for $V_1=0$ and $V_3=0.4t$ in \eqref{eq:H_kagome}. An interaction-driven TI appears for finite $U$ and $V_2$ and requires a critical interaction strength to appear, and some fine-tuning of parameters. Solid lines indicate first order and dashed lines second order transitions.  From Ref.[\onlinecite{Wen:prb10}].}
\label{fig:kagome13}
\end{figure}

\begin{figure}[th]
\centering
\includegraphics[width=0.9\linewidth]{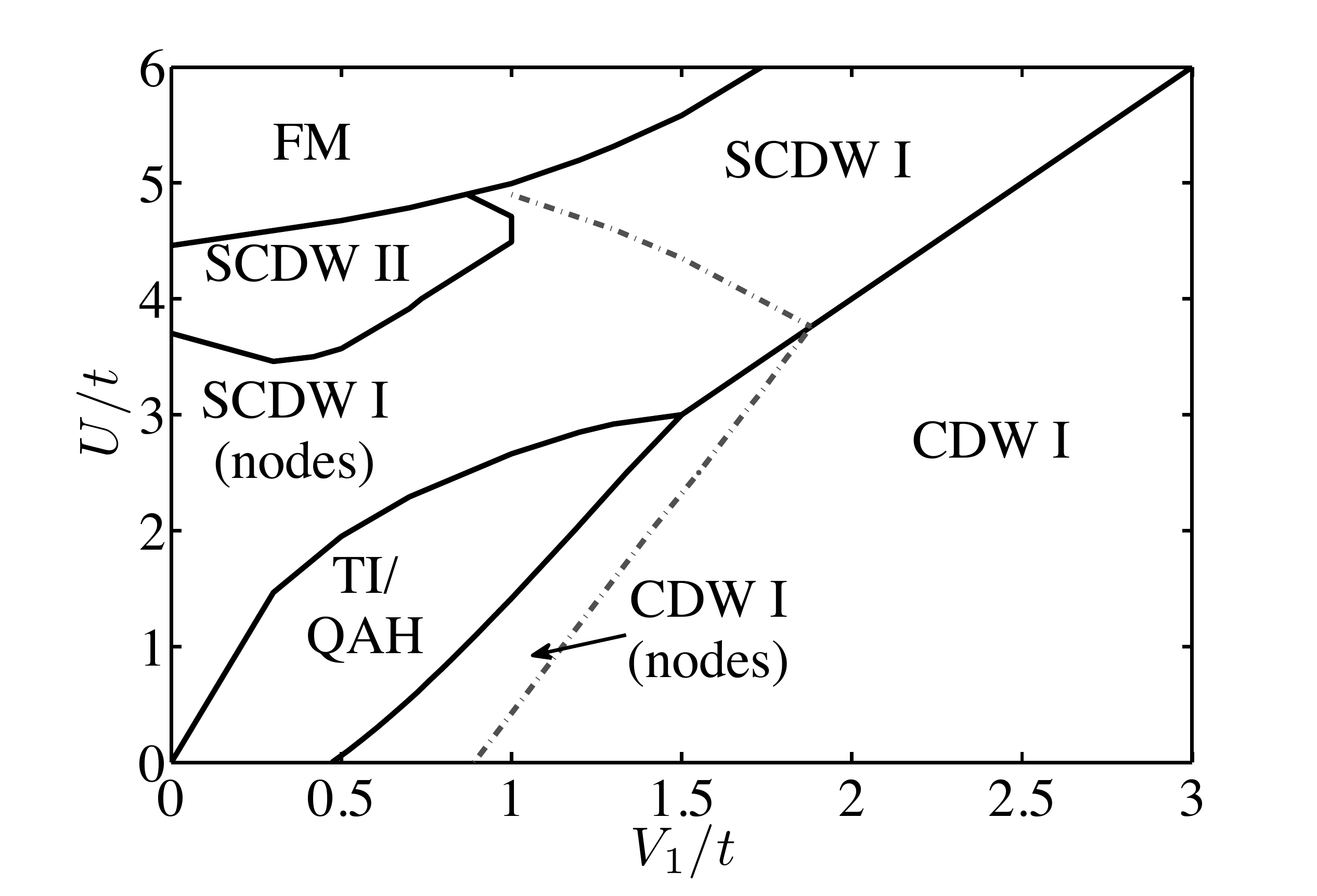}
\caption{The $U-V_1$ phase diagram for $V_2=V_3=0$ at $2/3$ filling fraction (quadratic band touching point) in \eqref{eq:H_kagome}.  Note that for small {\it generic} interactions the leading instability is a topological phase, either TI or the quantum anomalous Hall phase (QAH). From Ref.[\onlinecite{Wen:prb10}].}
\label{fig:kagome23}
\end{figure}

To perform the mean-field calculation, we decouple the on-site interaction according to\cite{Wen:prb10}
\begin{eqnarray}
&n_{i\uparrow}n_{i\downarrow}\approx n_{i\uparrow}\langle n_{i\downarrow}\rangle+\langle n_{i\uparrow}\rangle n_{i\downarrow}-\langle n_{i\uparrow}\rangle\langle n_{i\downarrow}\rangle\nonumber\\
&-c_{i\uparrow}^{\dag}c_{i\downarrow}\langle c_{i\downarrow}^{\dag}c_{i\uparrow}\rangle-\langle c_{i\uparrow}^{\dag}c_{i\downarrow}\rangle c_{i\downarrow}^{\dag}c_{i\uparrow}+\langle c_{i\uparrow}^{\dag}c_{i\downarrow}\rangle\langle c_{i\downarrow}^{\dag}c_{i\uparrow}
\rangle.
\end{eqnarray}
We assume the mean-field solutions are described by a colinear spin alignment and therefore, without loss of generality, we set $\langle c_{i\uparrow}^{\dag}c_{i\downarrow}\rangle=\langle c_{i\downarrow}^{\dag}c_{i\uparrow}\rangle=0$.\cite{Wen:prb10} The further-neighbor interaction is decoupled in a similar way:\cite{Wen:prb10}
\begin{eqnarray}
n_{i}n_{j}&\approx& n_{i}\langle n_{j}\rangle+\langle n_{i}\rangle n_{j}-\langle n_{i}\rangle\langle n_{j}\rangle-\sum_{\alpha\beta}\left(c_{i\alpha}^{\dag}c_{j\beta}\langle c_{j\beta}^{\dag}c_{i\alpha}\rangle\right.\nonumber\\
&&+\left.\langle c_{i\alpha}^{\dag}c_{j\beta}\rangle c_{j\beta}^{\dag}c_{i\alpha}-\langle c_{i\alpha}^{\dag}c_{j\beta}\rangle\langle c_{j\beta}^{\dag}c_{i\alpha}\rangle\right).
\end{eqnarray}
The results of the mean-field calculations are shown in Fig.~\ref{fig:kagome13} and \ref{fig:kagome23}.  For a detailed description of the phases appearing in the phase diagrams, see Ref.[\onlinecite{Wen:prb10}].  

The important message we would like to emphasize is that for Dirac points, Fig~\ref{fig:kagome13}, topological phases require a critical interaction strength that can be rather large, and appear to require some amount of ``fine tuning".\cite{Zhang:prb09}  For example, note that $V_1=0$ in Fig.~\ref{fig:kagome13}.
On the other hand, when quadratic band touching points are present in the non-interacting band structure, Fig.~\ref{fig:kagome23}, topological phases can appear as the leading instabilities, without any ``fine tuning" of the interactions.  Thus, two-dimensional band structures with flat band touching points but small (or even vanishing) spin-orbit coupling are excellent candidates for topological phases.\cite{FZhang:prl11} General arguments and functional renormalization group studies (that treat fluctuations) suggest the mean-field results may be reliable and that the TI phase is likely preferred over the quantum anomalous Hall (QAH) state when they are degenerate at the mean-field level.\cite{Raghu:prl08}

A final point on which we would like to remark concerning interacting topological insulators in two dimensions is how interactions may help identify the phase itself.   For example, if two edges of different quantum spin Hall systems are brought close to one another the edges will interact via the Coulomb interactions of the electrons in each edge.  Under some circumstances, this can lead to novel quasi-one dimensional phases.\cite{Tanaka:prl09}  
\begin{figure}[th]
\includegraphics[width=5cm]{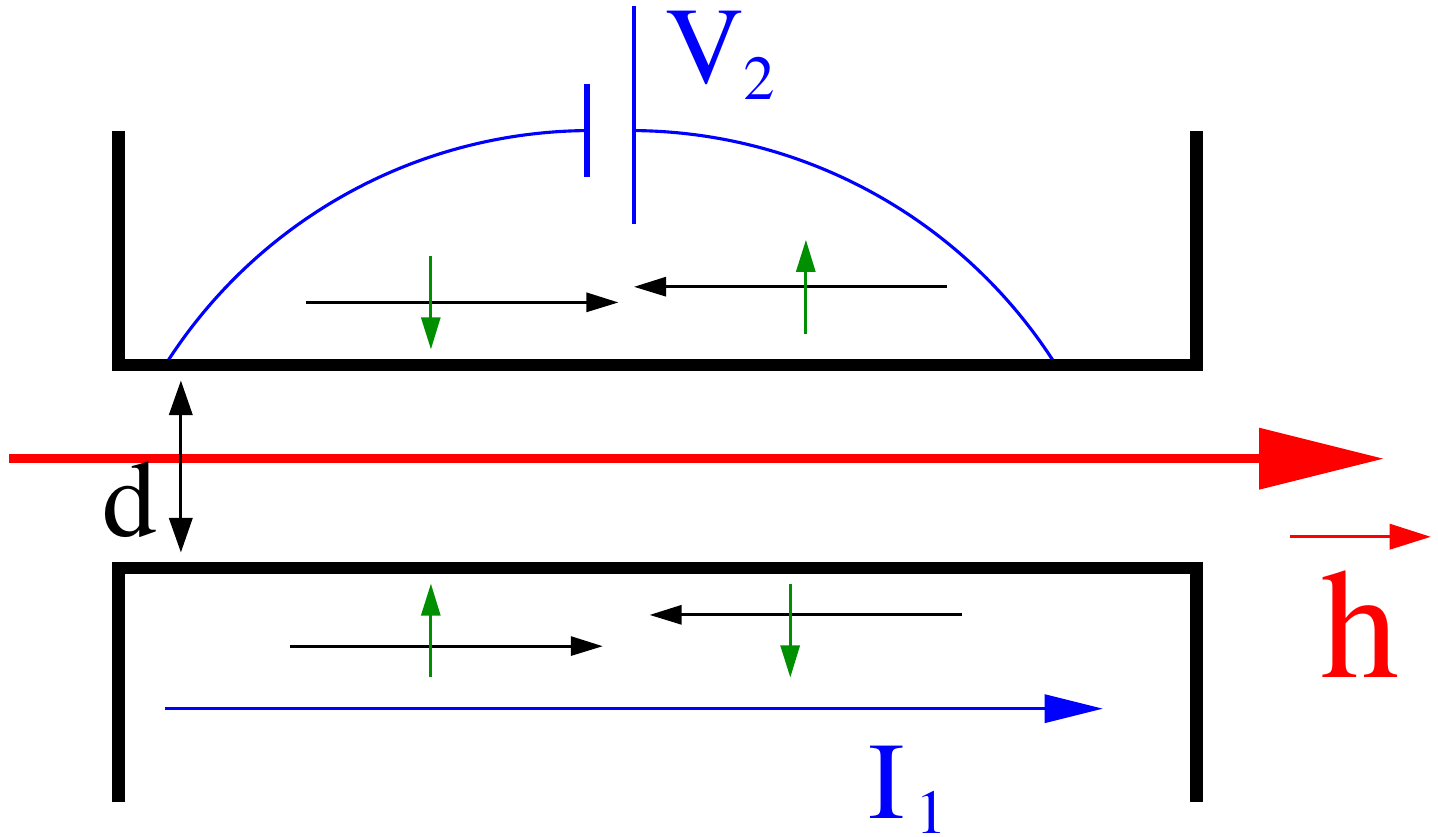}
\caption{(color online) Schematic of a drag measurement between two QSH systems from Ref.[\onlinecite{Zyuzin:prb10}].  A current $I_1$ is driven along the upper edge of the lower QSH system and through electron-electron interactions a voltage $V_2$ is induced in the lower edge of the upper QSH system. A magnetic field $\vec h$ is applied in the plane of wires, perpendicular to the spin quantization axis (assumed perpendicular to the plane of QSH systems).  Time-reversed Kramer's pairs are indicated for the two edges.  A QSH on top of QSH geometry could also be used.} \label{fig:drag}
\end{figure}
However, if the interactions are not too strong (the edges are far enough apart), a current driven in one edge will result in a voltage drop along the other edge in the so-called ``Coulomb drag" effect.\cite{Ponomarenko:prl00,Yakovenko:JETP92,Klesse:prb00,Fuchs:prb05,Pustilnik:prl03,Aristov:prb07,Fiete:prb06}  The properties of the edge are reflected in the temperature and magnetic field dependence of the drag. For identical edges described by a helical liquid,\cite{Wu:prl06,Xu:prb06} the Coulomb drag has a novel magnetic field dependence that is ``turned on" in the presence of a time-reversal symmetry breaking magnetic field,\cite{Zyuzin:prb10}
\begin{equation}
\label{eq:rD}
r_{D} \propto h^{4}T^{4K-3},
\end{equation}
where $h$ is the strength of the applied magnetic field and $K$ is the Luttinger parameter of the helical edge system.\cite{Zyuzin:prb10}  The role of the finite magnetic field is to open a back-scattering channel for electrons.  There is also a weak field dependence of the Luttinger parameter $K$.\cite{Zyuzin:prb10}

\subsection{Fractional topological insulators and flat band fractional quantum Hall effect }

The discussion in Sec. \ref{sec:non-interacting} which related the integer quantum Hall effect to TIs immediately raises the question of whether there is a ``fractional" analog to TIs that depends in an intrinsic way on electron-electron interactions in the same way the fractional quantum Hall effect does. In particular, is there a time-reversal invariant topological insulator that is not adiabatically connected to a free electron model? Theoretically, the answer is affirmative.\cite{Bernevig:prl06,Levin:prl09,Qi11}

In two dimensions, we are guided in a study of fractional topological insulators by the fractional quantum Hall effect.  If one has in hand a spinless fermion model that realizes a fractional quantum Hall effect, an $S^z$ conserving, time-reversal invariant model can be straightforwardly constructed by generalizing the Haldane-Kane-Mele correspondence to the interacting case.\cite{Bernevig:prl06,Levin:prl09} Only very recently has the explicit construction been understood for the non-$S^z$ conserving case (in the presence of Rashba spin-orbit coupling, for example).\cite{Qi11}

A key to understanding lattice models of fractional topological insulators in two dimensions is thus understanding the fractional quantum Hall effect in lattice models. (While this appears necessary, it is not sufficient for constructing a general many-body wavefunction.\cite{Wang:prl11,Sun:prl11,Neupert:prl11,Tang:prl11,Qi11})  For a while, this understanding eluded the community whose picture was based on trial wavefunctions and effective low-energy theories,\cite{Nayak:rmp08} but in the past year important progress has been made.  

Attention has focused on a class of insulators with nearly flat bands that possess finite Chern numbers.   With a finite Chern number, a partial filling of the flat bands can lead to a fractional quantum Hall effect because interaction energy will always dominate kinetic energy.\cite{Sheng11}  To date, only a few lattice models have been proposed which are expected to lead to a fractional quantum Hall effect.\cite{Wang:prl11,Sun:prl11,Neupert:prl11,Tang:prl11,Hu:prb11}  The relevant figure of merit in such models is the ratio of the band gap to the bandwidth of the flat band with a finite Chern number.  In the model we discuss below, we find this ratio can be as high as 70, which is among the largest in the models reported in the literature thus far.\cite{Wang:prl11,Sun:prl11,Neupert:prl11,Tang:prl11,Hu:prb11}

\begin{figure}[t]
\includegraphics[width=0.8\linewidth]{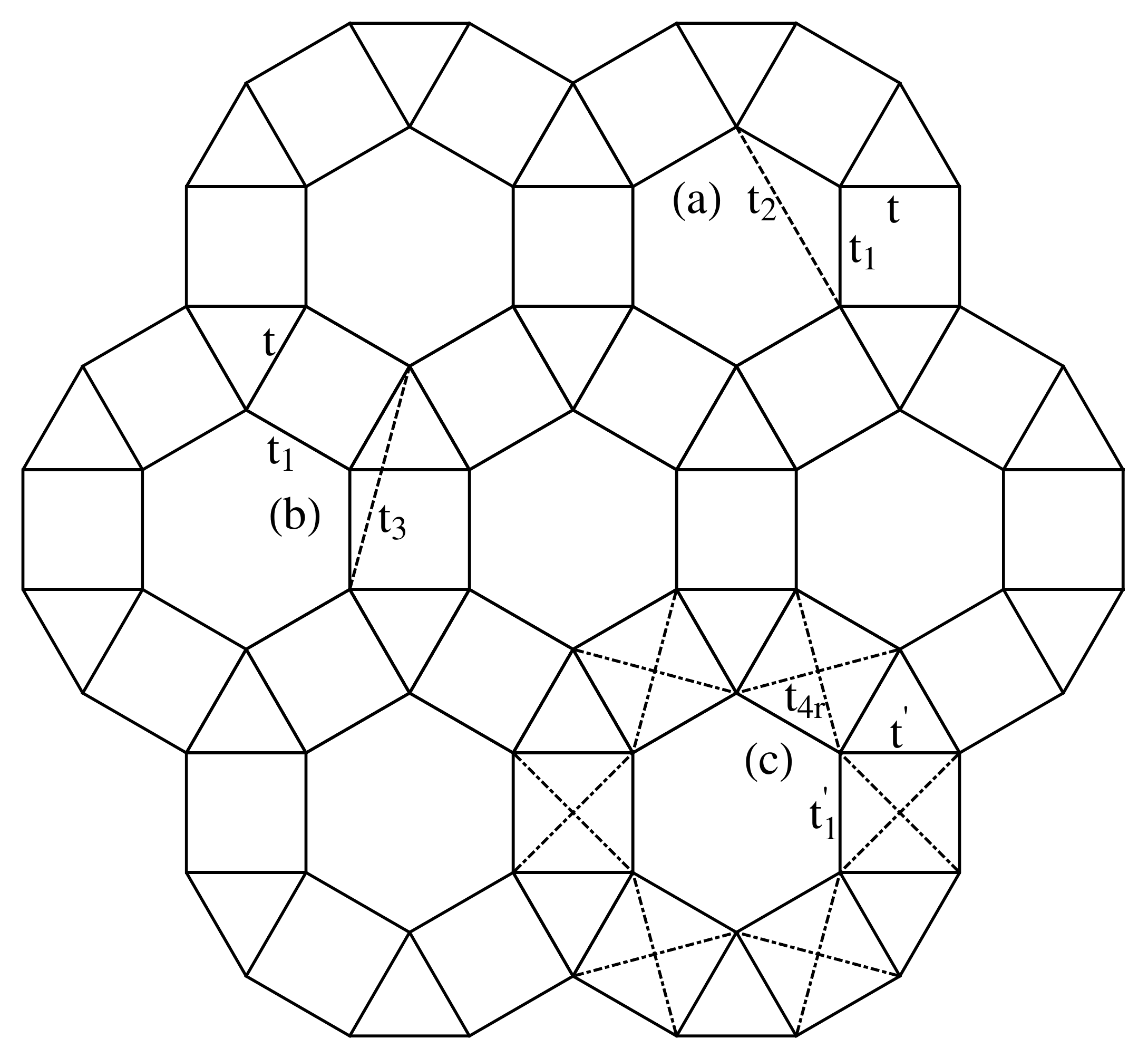}
\caption{Schematic of the ruby lattice and illustration of the nearest-neighbor hopping, $t,t_1$ (real) and $t',t_1'$ (complex), and the three types of ``second-neighbor" spin-orbit coupling or hopping indicated by the dashed or dot dashed lines, $t_2,t_3$ and $t_{4r}$ (from Ref.[\onlinecite{Hu:prb11}]).  (a) The spin-orbit coupling strength $t_2$ within a hexagon. (b) The spin-orbit coupling strength $t_3$ within a pentagon composed of one triangle and one square.  The hoppings $t_2$ and $t_3$ are present on all such bonds of the type shown that are consistent with the symmetry of the lattice. (c) Schematic of the hopping parameters used to obtain a flat band with a finite Chern number and $W/E_g \approx 70$.}\label{fig:ruby}
\end{figure}

We study\cite{Hu:prb11} a simple $s$-band tight-binding model on the ruby lattice shown in Fig.~\ref{fig:ruby}.
For fermions with spin, the Hamiltonian is given by ${\cal H}={\cal H}_{\rm ruby}+{\cal H}_{\rm ruby-SO},$ where
\begin{equation}
\label{eq:H0_ruby}
{\cal H}_{\rm ruby}=-t\sum_{i,j\in\triangle, \sigma}c_{i\sigma}^\dag c_{j\sigma}
    -t_1\sum_{\triangle\rightarrow\triangle,\sigma}c_{i\sigma}^\dag
    c_{j\sigma},
\end{equation}
and
\begin{equation}
\label{eq:H_SO_ruby}
{\cal H}_{\rm ruby-SO}=it_2\!\!\!\!\sum_{\ll ij\gg, \alpha\beta}\!\!\!\!\nu_{ij}s^z_{\alpha\beta}
c_{i\alpha}^\dag c_{j\beta}+it_3\!\!\!\!\sum_{\ll ij\gg, \alpha\beta}\!\!\!\!\nu_{ij}
s^z_{\alpha\beta}c^\dagger_{i\alpha}c_{j\beta}
\end{equation}
on the ruby lattice shown in Fig.~\ref{fig:ruby}. Here $c_{i\sigma}^\dag/c_{i\sigma}$ is the creation/annihilation operator of an electron on site
$i$ with spin $\sigma$. As indicated in Fig.~\ref{fig:ruby}(a,b), $t$ and $t_1$ are real first-neighbor hopping parameters, and $t_2$, $t_3$ are real second-neighbor hoppings (these appear with the imaginary number $i$ in Eq.\eqref{eq:H_SO_ruby} making the total second-neighbor hopping purely imaginary and time-reversal symmetric). The quantity  $\nu_{ij}$ is equal to 1 if the electron makes a left turn on the lattice links during the second-neighbor hopping, and is equal to -1 if the electron make a right turn during that process. As is clear from Fig.\ref{fig:ruby}, the unit cell of the lattice contains six sites, so six two-fold degenerate bands will result.  In addition to the real hopping parameters $t,t_1$ in \eqref{eq:H0_ruby}, symmetry also allows complex, spin-dependent nearest neighbor hopping with imaginary components $t',t_1'$, as shown in Fig.~\ref{fig:ruby}. These parameters lead to a variety of time-reversal invariant phases (including TI) and a rather complex phase diagram.\cite{Hu:prb11}

\begin{figure}[th]
\centering
\includegraphics[width=0.9\linewidth]{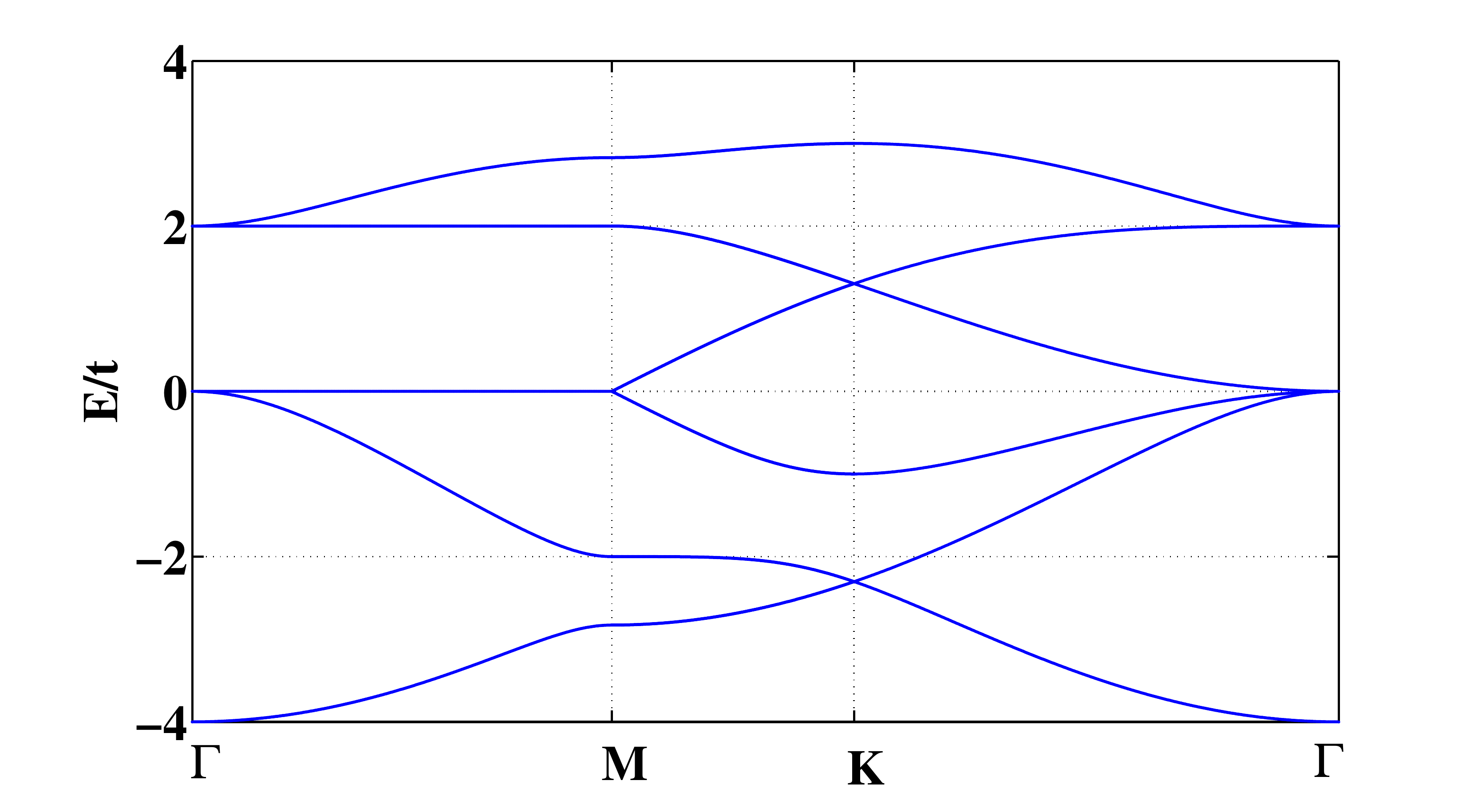}\\
(a)\\
\includegraphics[width=0.9\linewidth]{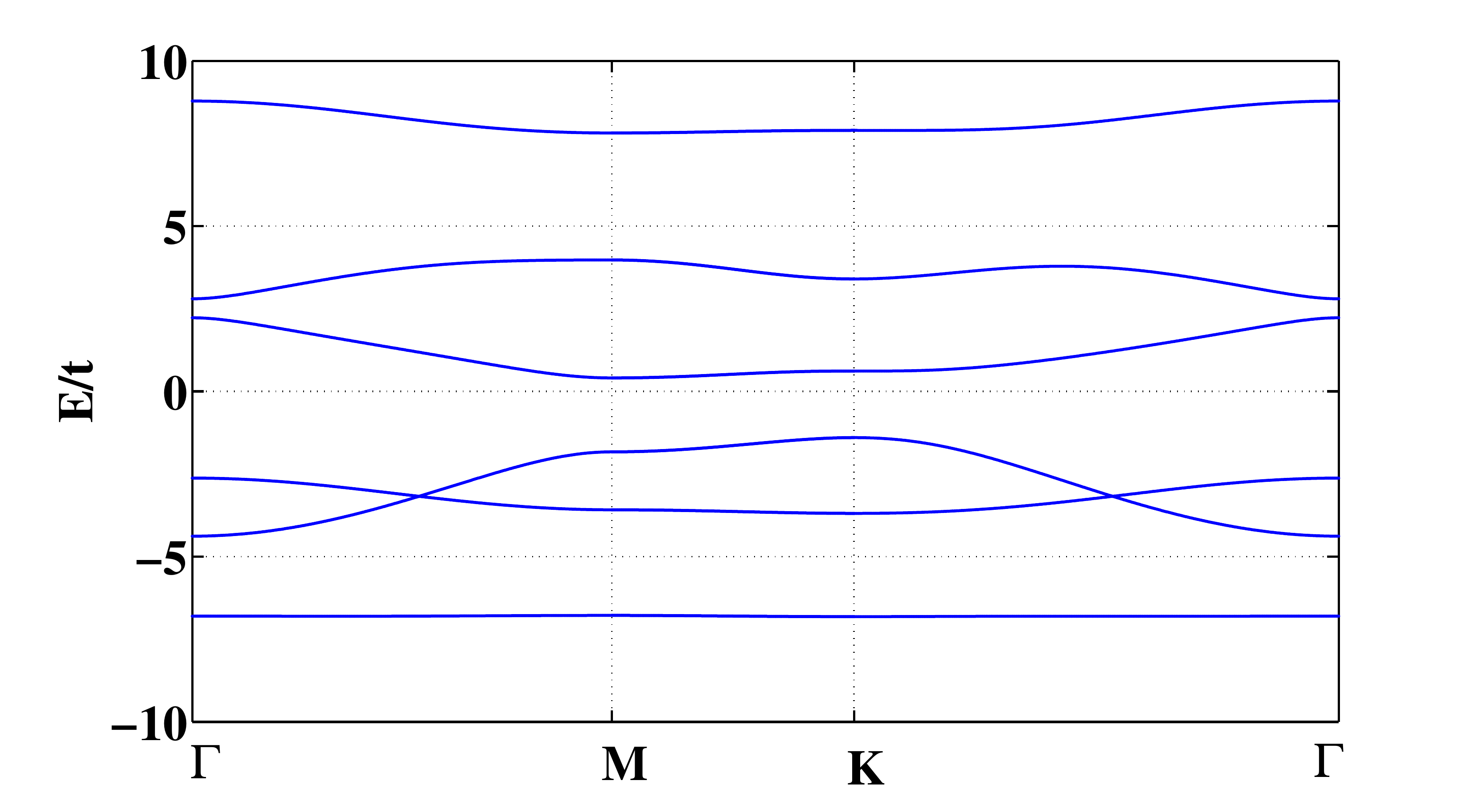}\\
(b)
\caption{(a) The energy bands without any spin-orbit coupling and with it tuned on, given by Eq.\eqref{eq:H0_ruby} with $t=t'$ real along high symmetry directions. Note that there is a Dirac point at $K$ for 1/6 and 2/3 filling, and a quadratic band touching point at $\Gamma$ for 1/2 and 5/6 filling. (The underlying lattice is triangular, as it is for the honeycomb lattice.) Note also the flat bands along the $\Gamma-M$ direction at 1/2 and 5/6 filling.  (b) The energy bands with $t_i=1.2t, t_{1r}=-1.2t, t_{1i}=2.6t, t_{4r}=-1.2t$ in \eqref{eq:t1complex}. The lowest energy band has Chern number -1. The ratio of the band-width to band gap is approximately 70 (from Ref.[\onlinecite{Hu:prb11}]).}\label{fig:bands}
\end{figure}

Here we are mainly interested in the spinless case with broken time-reversal symmetry and the hopping parameters described in Fig.~\ref{fig:ruby}(c). For spinless fermions we study \eqref{eq:H0_ruby} with
\begin{eqnarray}
t'&=&t+\text{i}\sigma_z t_i,\nonumber\\
t_1'&=&t_{1r}+\text{i}\sigma_z t_{1i},
\label{eq:t1complex}
\end{eqnarray}
and add the hopping inside the square in the diagonal directions, labelled $t_{4r}$.  This is shown schematically in Fig.~\ref{fig:ruby}(c). A nonzero Chern number and nearly flat band occurs in this case. For example, with $t_i=1.2t, t_{1r}=-1.2t, t_{1i}=2.6t, t_{4r}=-1.2t$ the Chern number is -1 and the $\text{gap}=2.398t, \text{and band width}=0.037t$, which gives $W/E_g \approx 70$, and therefore makes an excellent candidate to realize a fractional topological insulator.\cite{Hu:prb11}  The corresponding band structure is shown in Fig.~\ref{fig:bands}(b).  The spinless band structure for \eqref{eq:H0_ruby} with $t'=t=1$ is shown in  Fig.~\ref{fig:bands}(a).  Note the interesting band crossing/touching points and flat portions of the bands even in this case.

A new direction has emerged in the search for systems with flat bands with conditions favorable to realizing fractional topological insulators or fractional quantum anomalous Hall states--transition metal oxide interfaces.\cite{Xiao11}  Research in this direction is rapidly evolving with a variety of interesting interaction-driven phases possible.\cite{Ruegg11_2,Wang11,Yang11}

As rich as the physics of topological insulators and quantum spin liquids is in two dimensions, the behavior is even richer in three dimensions.  This is because the three-dimensional time-reversal invariant topological insulators are qualitatively different from their two-dimensional counter parts, the latter of which can always be viewed as being derived from an underlying quantum Hall effect.  The so-called ``strong topological" insulators\cite{Fu:prl07} (non-interacting) are an entirely distinct three-dimensional phase with novel response properties.\cite{Qi:prb08,Qi:sci09,Essin:prl09,Essin:prb10,Fu:prl07,Teo:prb08,Rosenberg:prb10,Rosenberg_2:prb10} Moreover, once interactions are brought into the picture the physics is richer still. We now turn to three dimensions.

\section{Three-dimensional systems}
\label{sec:3D}

One of the salient features of TIs is that their boundaries possess a topologically protected ``metallic" state that is robust to disorder.\cite{Schnyder:prb08,Roy_2:prb09,Roy:prb09,Moore:prb07,Fu:prl07,Wu:prl06,Xu:prb06}  Both the two dimensional\cite{Strom:prl09,Huo:prl09,Xu:prb10,Teo:prb09,Law:prb10,Zyuzin:prb10,Tanaka:prl09,Maciejko:prl09} and the three dimensional\cite{Qi:prb08,Qi:sci09,Essin:prl09,Essin:prb10} boundaries have been shown to exhibit interesting responses to perturbations. In the experimental literature on TIs, the three dimensional systems have dominated. The weakly-interacting nature of topological insulators has enabled accurate predictions based on density functional theory\cite{Bernevig:sci06,Fu:prb07,Teo:prb08,Zhang:np09} for a wide range of two and three dimensional systems,\cite{Chadov:nm10,Wang:prl11,Feng:prl11,Zhang:prl11,Xiao:prl10,Lin:prl10,Chen:prl10,Yan:epl10,Yan:prb10,Lin:cm10,Yan:11,Lin:nm10,Sun:prl10,Lin:prl10,Feng:prb10,Al-Sawai:prb10} and experiment has followed with confirming data in a large and rapidly growing number of instances.\cite{Konig:sci07,Roth:sci09,Hsieh:nat08,Hsieh:sci09,Xia:np09,Chen:sci09,Hsieh:nat09,Hsieh:prl09,Hor:prb09,Sato:prl10,Kuroda:prl10,Nishide:prb10} There are also theoretical predictions for systems closely related to TIs with Weyl fermions in their bulk and unusual boundary excitations,\cite{Wan:prb11,Wan11,Burkov:prl11,Burkov11,Yang11, Halasz11} as well as crystalline topological insulators.\cite{Fu:prl11}

In this section we will describe systems that are too strongly interacting to allow  a naive application of density functional theory.  On the most exotic side, three-dimensional versions of exotic fractional topological insulators with a non-trivial ground state degeneracy have been proposed.\cite{Maciejko:prl10,Swingle:prb11,Cho:ap11,Levin:arxiv11} For interacting systems treated at the mean-field level, a spontaneous generation of spin-orbit coupling (with topological defects possible) can occur, similar to two dimensions.\cite{Zhang:prb09}  

Here, we will focus on semi-realistic models that are motivated by the physics of transition metal oxides, famous in recent years for high-temperature superconductivity\cite{Lee:rmp06,Dagotto:rmp94}  and colossal magnetoresistance.\cite{Salamon:rmp01}  These materials typically involve $3d$ orbitals on the transition metals and lead to strong Coulomb interactions due to their localized nature (compared to $4d$ and $5d$ orbitals).  However, $4d$ and $5d$ orbitals in layered perovskites such as $\mathrm{Sr_2RuO_4}$, $\mathrm{Sr_2RhO_4}$, $\mathrm{Sr_2IrO_4}$, $\mathrm{Na_2IrO_3}$, and the hyperkagome $\mathrm{Na_4Ir_3O_8}$ are more spatially extended and thus the Coulomb interaction is typically weaker than those with $3d$ orbitals.\cite{ryden:prb70} The hyperkagome $\mathrm{Na_4Ir_3O_8}$ is an important three-dimensional quantum spin liquid candidate.\cite{Okamoto:prl07,Zhou:prl08,Lawler:prl08,Podolsky:prb11,podolsky:prl09} 

\begin{figure}[t]
\begin{center}
\includegraphics[width=8cm]{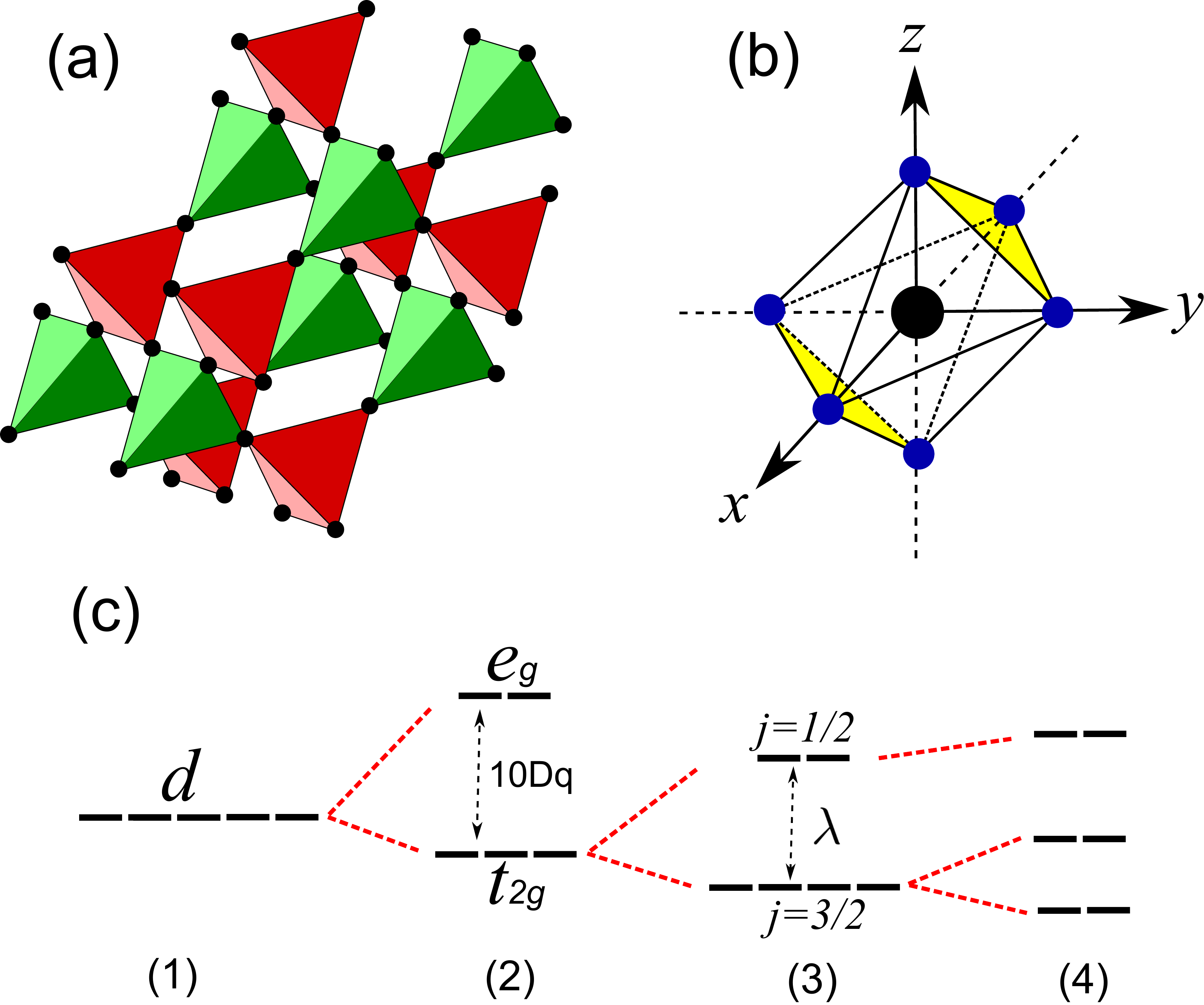}
\caption{(Color online) From Ref.[\onlinecite{Kargarian:prb11}]. (a) An illustration of the pyrochlore lattice which is composed of corner sharing tetrahedra. Transition elements are indicated by black solid circles. (b) Each transition ion is surrounded by an oxygen octahedron shown by six solid blue (dark grey) circles. A transition ion is located at the origin of the local coordinate and is shown in black. We study a trigonal distortion  preserving $C_3$ symmetry applied along the [111] direction (or its equivalent), shown by two yellow (grey) faces, and an elongation preserving $C_4$ symmetry along the $z$-axis of the local coordinate. (c) A schematic representation of the splitting of the bare atomic $d$-levels (1), due to a cubic crystal field arising from the octahedral environment (2), unquenched spin-orbit coupling in the $t_{2g}$ manifold (3), and a distortion of the octahedron (4). The values of the splittings in (4) depend on $\lambda$ and $\Delta_{3,4}$.} \label{pyrochlore}
\end{center}
\end{figure}

The more extended nature of the $4d$ and $5d$ orbitals compared to the $3d$ orbitals leads to a greater level splitting in a crystal field, and enhances their sensitivity to lattice distortions.\cite{Yang:prb10} In many oxides, the transition ions are surrounded by an octahedron of oxygen atoms, $\mathrm{MO_6}$, where M represents a transition metal ion. The crystal field splits the 5 degenerate (neglecting spin for the moment) $d$-orbitals into two manifolds (see Fig.\ref{pyrochlore}c): a lower lying $t_{2g}$ ($d_{xy},d_{yz},d_{zx}$) manifold and a higher lying $e_{g}$ ($d_{3z^2-r^2},d_{x^2-y^2}$) manifold.\cite{Maekawa,Tokura_2000} The energy separation between the $t_{2g}$ and $e_{g}$ levels is conventionally denoted ``10Dq" and is typically on the order of  $\sim$1-4 eV, which is large compared to many $3d$ compounds.\cite{moon:prb06}

Besides the crystal field, the relativistic spin-orbit coupling is another energy scale that results from the large atomic numbers of heavy transition elements. While in the absence of spin-orbit coupling the on-site Coulomb interaction is of the same order as the band width,\cite{Nakatsuji:prl00,lee:prb01} inclusion of strong spin-orbit coupling modifies the relative energy scales.\cite{Pesin:np10} Thus, for materials with $4d$ and particularly $5d$ electrons, one expects the appearance of novel phases with unconventional electronic structure and order due to the characteristic energy of spin-orbit coupling approaching that of the Coulomb interactions.\cite{Pesin:np10,Chen:prb10,Kargarian:prb11}

In this section we focus on the interplay and competition between strong correlation effects, spin-orbit coupling, and lattice distortion that is expected to be important in heavy transition metal oxides with a pyrochlore lattice.\cite{Kargarian:prb11}   In the heavy transition metal oxides one expects both the spin-orbit coupling\cite{shitade:prl09,Dodds:prb11} and the lattice distortion energies\cite{Landron:prb08,Yang:prb10} to be of the order of $0.05-0.5$ eV, while the interaction energy is typically at the higher end of this scale to somewhat larger, $0.5-2$ eV.\cite{shitade:prl09,Wan:prb11,Wan11}  While the phase diagram of an interacting undistorted pyrochlore model with $j=1/2$ has already been studied,\cite{Pesin:np10} we expand those results to include the effects of distortions of the local octahedra on the phase diagram.\cite{Kargarian:prb11}  

We also investigate pyrochlore oxides at different $d$-level fillings with the Fermi energy lying in the quadruplet $j=3/2$ manifold, which has not been considered in previous works.\cite{Kargarian:prb11}  One of our motivations is to see if the $j=3/2$ manifold can also realize the interesting Mott phases of the $j=1/2$ manifold.\cite{Pesin:np10}  We find that, indeed, these exotic phases can be realized for the $j=3/2$ manifold.  Moreover, we find that for the $j=1/2$ manifold ``weak" topological variants of the exotic Mott phases can also appear in the phase diagram when certain types of lattice distortion are present.

To study the effects of lattice deformations,\cite{Bergman:prb06,Yang:prb10} we assume that the octahedron surrounding an ion can be distorted in two ways: (1) a trigonal distortion preserving local $C_3$ symmetry and (2) an elongation (expansion) of octahedra preserving local $C_4$ symmetry. (See Fig.\ref{pyrochlore}b.) The former has been argued to be rather common and can be described by the following Hamiltonian on each transition metal ion site:\cite{Yang:prb10} 
\begin{equation} 
\label{c3}
{\cal H}_{tri}=-\Delta_3 (d_{yz}^{\dagger}d_{zx}+d_{yz}^{\dagger}d_{xy}+d_{zx}^{\dagger}d_{xy})+h.c.,
\end{equation}
where $\Delta_3$ parameterizes the strength and sign of the $C_3$ preserving distortion, and the $C_4$ elongation/contraction splitting is described by\cite{Chen:prb10} 
\begin{equation}  
\label{c4} 
{\cal H}_{el}=\Delta_4 l_{z}^{2}=\Delta_4(n_{yz}+n_{zx}), 
\end{equation} 
where $\Delta_4$ parametrizes the strength and sign of the distortion, and $l_z$ is the $z$ component of the effective angular momentum of the $t_{2g}$ orbitals related to the occupation of the $d_{xy}$ orbital by $n_{xy}=n_d-(l_z)^2$ which follows from the constraint $n_d=n_{xy}+n_{yz}+n_{zx}$.\cite{Chen:prb10} For an elongation of the tetrahedron, $\Delta_4<0$, and for a compression of the tetrahedron, $\Delta_4>0$.  Trigonal distortions appear to be more common in real materials, and the magnitude of the energy splittings can be crudely estimated from density functional theory calculations based on X-ray determined positions of oxygen atoms around the transition metals.  We are not aware of detailed calculations of this type for the 4$d$ and 5$d$ pyrochlore oxides, but closely related 3$d$ systems appear to have splittings on the level of $0.01-0.5$ eV.\cite{Landron:prb08}  We take this as crude estimate, with the larger end of the energy scale probably more likely for the more extended 4$d$ and 5$d$ orbitals.

Thus, the local Hamiltonian describing the $t_{2g}$ orbitals on each site is
 \begin{equation} 
 \label{local}  
{\cal H}_{local}={\cal H}_{so}+{\cal H}_{tri}+{\cal H}_{el},
 \end{equation}
with 
\begin{equation} \label{local_so}
{\cal H}_{so}=-\lambda l\cdot s,
\end{equation}
where $l=1$ and $s=1/2$ describe the orbital and spin degrees of freedom, and $\lambda>0$ parameterizes the strength of the spin-orbit coupling. That the $t_{2g}$ orbitals can be effectively described by angular momentum $l=1$ comes from the projection of the $d$-orbital angular momentum into the local basis of $t_{2g}$ manifold.\cite{Pesin:np10,Chen:prb10}

The Hamiltonian \eqref{local} can be easily diagonalized and its eigenvectors describe a projection onto the spin-orbit plus distortion basis. We will denote the projection by a matrix $M$, which contains all the information about the spin-orbit coupling and the distortion of the octahedra (all assumed identical so translational invariance is preserved). Moreover, due to the presence of time-reversal symmetry, the eigenvectors form Kramers pairs. A schematic representation of splitting $t_{2g}$ upon including the terms in Eq.\eqref{local} is shown in Fig.\ref{pyrochlore}c.

We now turn to a derivation of the effective Hamiltonian. We first assume $\lambda=\Delta_3=\Delta_4=0$, {\em i.e.} neglect the contributions in \eqref{local}. To obtain the kinetic terms of the Hamiltonian, we need to describe the $t_{2g}$ orbitals of a single ion in the local coordinate system defined by the octahedron of oxygen atoms surrounded the ion, and we need the $p$-orbitals of oxygen in the global coordinate system.\cite{Pesin:np10}  The hopping of electrons from one transition metal ion to a nearest-neighbor transition metal ion is mediated by the oxygen $p$-orbitals.  (We note that for the relatively extended $5d$ orbitals direct overlap may also be important, as well as further neighbor hopping.\cite{jin:arxiv09}) We thus compute the $p$-$d$ overlaps to determine the hopping matrix elements. The local and global axes are related by a set of rotation matrices.\cite{Pesin:np10,Yang:prb10} The combination of rotation matrices and $d$-$p$ overlaps gives rise to the following Hamiltonian:
\begin{equation} 
\label{H_d}
{\cal H}_{d}=\varepsilon_d \sum_{i\gamma\sigma}d_{i\gamma\sigma}^{\dag}d_{i\gamma\sigma}+t\sum_{<i\gamma \sigma,i'\gamma' \sigma'>}T^{ii'}_{\gamma \sigma,\gamma' \sigma'}d_{i\gamma\sigma}^{\dag}d_{i'\gamma'\sigma'},
 \end{equation} 
 where $i$, $\gamma$, and $\sigma$ in the sums run over lattice sites, $t_{2g}$ orbitals ($xy,yz,zx$), and spin degrees of freedom, respectively. The $\varepsilon_{d}$ stands for the onsite energy of the degenerate $t_{2g}$ orbitals, and $t=\frac{V^{2}_{pd\pi}}{\varepsilon_p-\varepsilon_d}$ is the unrotated hopping amplitude depending on the overlap integral $V_{pd\pi}$ and the energy difference between $p$ and $d$ orbitals. The parameter $t$ sets the basic hopping energy scale in the problem. Without loss of generality we set $\varepsilon_d=0$. 
 
 \begin{figure}[h]
\begin{center}
\includegraphics[width=8cm]{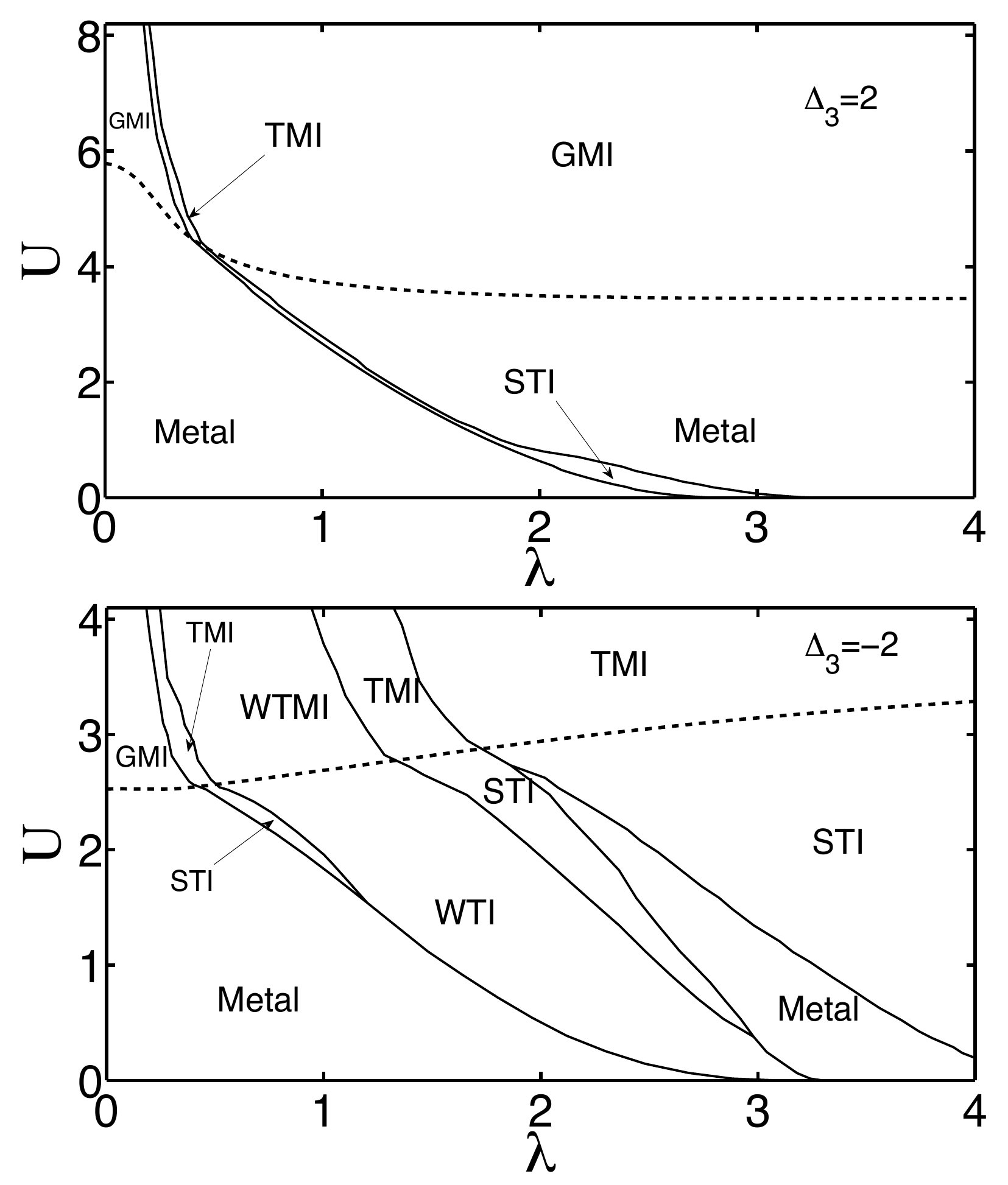}
\caption{Phase diagram of the $j=1/2$-band model corresponding to $n_{d}=5$ with positive $\Delta_3=2t$ (upper panel) and negative trigonal distortion $\Delta_3=-2t$ (lower panel). The dashed line separates the rotor condensed phases (below) from the uncondensed phases (above). We set $t=1$, and the phases labeled are as follows: Strong topological insulator (STI), Weak topological insulator (WTI), Gappless Mott insulator (GMI), Topological Mott insulator (TMI), Weak topological Mott insulator (WTMI) and Metallic phases. From Ref.[\onlinecite{Kargarian:prb11}].} \label{j12_ph}
\end{center}
\end{figure}

The effect of spin-orbit coupling and distortion are included via the projection of Hamiltonian in Eq.\eqref{H_d} into the eigenvectors of the local Hamiltonian in Eq.\eqref{local} using matrix $M$ as follows: 
 \begin{equation} 
 \label{H0} 
 {\cal H}_{\rm pyro}=\sum_{i\alpha}\upsilon_{\alpha}c^{\dag}_{i\alpha}c_{i\alpha}+t\sum_{<i\alpha,i'\alpha'>}\Gamma^{ii'}_{\alpha,\alpha'}c^{\dag}_{i\alpha}c_{i'\alpha'},
 \end{equation} 
 where $\upsilon_{\alpha}$ ($\alpha=1,...,6$) stands for the six eigenvalues of local Hamiltonian \eqref{local}, and the matrix $\Gamma$ describes the hopping between sites given in the local basis via $\Gamma=M^{*}TM^{T}$. The $c^{\dag}_{i\alpha}(c_{i\alpha})$ is the creation (annihilation) operator of an electron at site $i$ and in local state $\alpha$. Finally, we add a Coulomb interaction to obtain 
 \begin{equation}
 \label{hubbard}
{\cal H}={\cal H}_{\rm pyro}+\frac{U}{2}\sum_{i}(\sum_{\alpha}c^{\dag}_{i\alpha}c_{i\alpha}-n_d)^2,
 \end{equation} 
 where $U$ is the on-site Coulomb interaction and $n_d$ is the number of electrons on the $5d$ orbital of the transition metal ion. In the remainder of this section, we investigate the zero-temperature phase diagram of the full Hamiltonian \eqref{hubbard}, which includes the spin-orbit coupling and lattice distortions in \eqref{local}.  A full description of our results is given in Ref.[\onlinecite{Kargarian:prb11}].

\begin{figure}[h]
\begin{center}
\includegraphics[width=7.7cm]{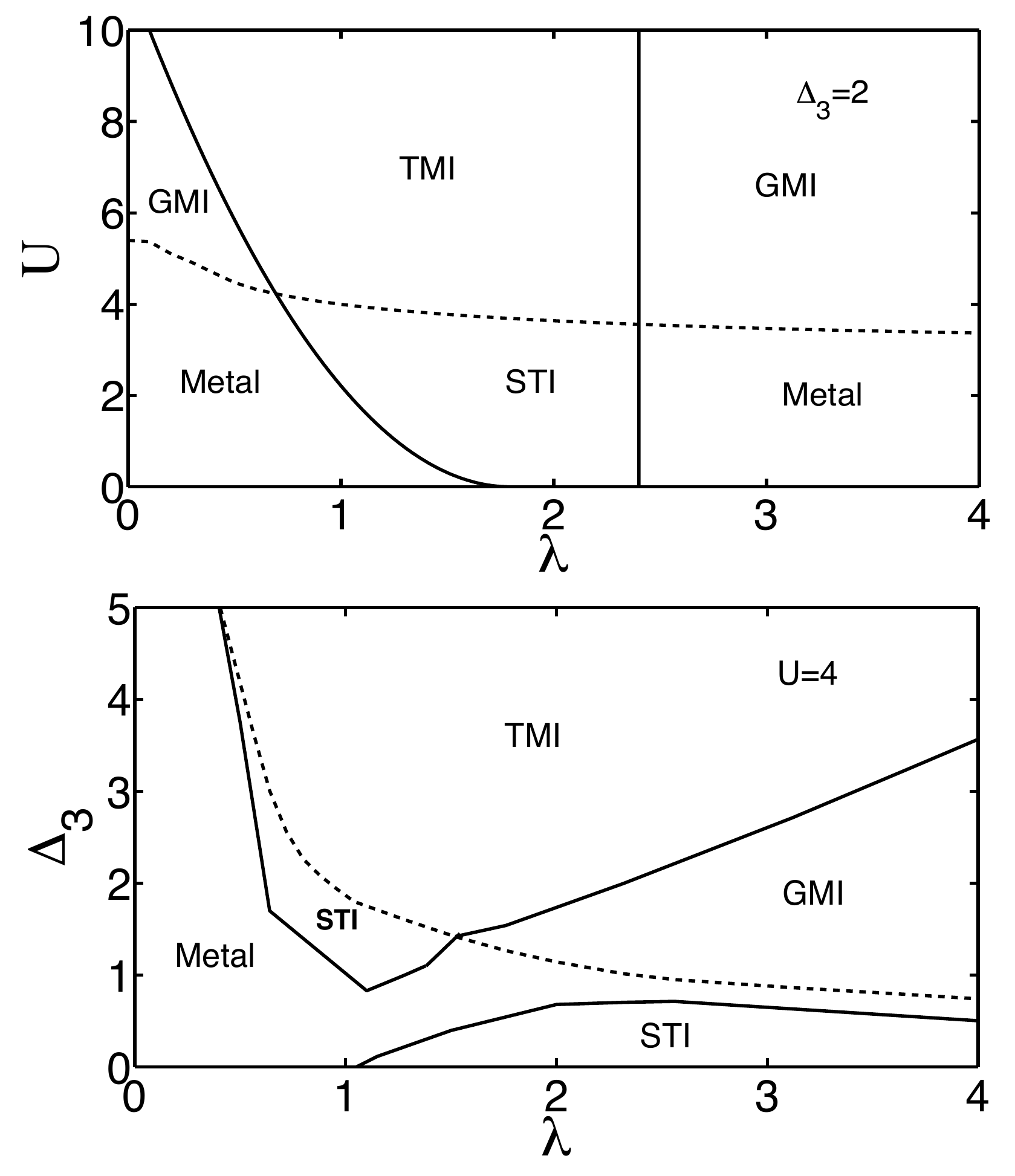}
\caption{Phase diagram of the $j=3/2$ band model with $n_d=3$, including the trigonal distortion of the octahedra. The labeling of the phases is the same as that used in Fig.\ref{j12_ph}. In the upper panel  $\Delta_3=2t$.  In the lower panel the interaction is fixed at $U=4t$ and the strength $\Delta_3$ of the trigonal distortion is varied, illustrating possible phases that may arise upon the application of pressure to a real system.  All energies are expressed in units of $t$. In both phase diagrams the dashed line separates the rotor uncondensed phase (above) from the condensed phase (below). We note the ``pocket" of STI around $\lambda \approx 1,\Delta_3\approx 1.5$ in the lower figure has a numerically difficult to determine boundary with the metallic phase; we have present our best assessment. From Ref.[\onlinecite{Kargarian:prb11}]. } \label{j32_c3}
\end{center}
\end{figure}

We apply the slave-rotor mean-field theory developed by Florens and Georges\cite{florens:prb02,florens:prb04} to treat the effect of weak to intermediate strength Coulomb interactions in the regime where the charge fluctuations remain important. In this theory each electron operator is represented in terms of a collective phase, conjugate to charge, called a rotor and an auxiliary fermion called a spinon as
\begin{equation} 
\label{slave_rotor} 
c_{i\alpha}=e^{i\theta_i}f_{i\alpha}, 
\end{equation} 
where $c_{i\alpha}$ is the electron destruction operator at site $i$ with quantum number $\alpha$, representing the states in \eqref{hubbard}. The factor $e^{i\theta_i}$ acting on the charge sector is a rotor lowering operator (with $\theta_i$ a bosonic field), and $f_{i\alpha}$ is the fermionic spinon operator.  The product of the two results in an object with fermi statistics, needed for the electron. Note the rotor part only carries the charge degree of freedom while the spinon part carries the remaining degrees of freedom $\alpha$. Therefore, an electron has natural spin-charge separation if $\alpha$ is spin in this representation.  Substituting \eqref{slave_rotor} into \eqref{hubbard} and applying mean-field theory results in the phase diagrams shown in Figs.\ref{j12_ph} and \ref{j32_c3}.\cite{Kargarian:prb11}  The two to three dimensional crossover and issues associated with gauge fluctuations beyond mean-field theory have been discussed in Ref.[\onlinecite{Witczak-Krempa:prb10}].  The method correctly recovers the $U\to 0$ limit in mean-field theory.

There are several features that are important to emphasize.  First, we have identified a new phase--the weak topological Mott insulator (WTMI)--in the regions shown in  Fig.~\ref{j12_ph}(b).\cite{Kargarian:prb11}  This phase is expected to have gapless thermal transport {\em without charge transport} along a certain class of topological defects, as described in Ref.[\onlinecite{Ran:np09}]. Also, we expect a rather rich phenomenology in the spin sector of WTMIs, similar to that recently emphasized in weak topological insulators (WTI).\cite{Ringel11} Second, it is clear from Fig.~\ref{j12_ph} that the sign of the trigonal distortion makes a big difference in the complexity of the phase diagram, with the negative trigonal distortion producing a more interesting phase diagram.  Third, comparing Figs. \ref{j12_ph} and \ref{j32_c3} one sees that the $j=1/2$ and $j=3/2$ manifolds behave differently for the same distortion.
Fourth, the lower panel of Fig. \ref{j32_c3} shows the phase diagram in the distortion-spin-orbit coupling plane.  The distortion axis can be viewed as a pressure axis.  Thus, the application of pressure to transitional metal oxides can drive the system through a complex set of phases, some of which are topological.  This is potentially a useful experimental parameter when searching for some of the exotic phases predicted here.

Our slave rotor results make a natural connection to quantum spin liquids in three dimensions described by fermionic excitations:  {\em All the phases above the dashed lines in Figs.\ref{j12_ph} and \ref{j32_c3} are time-reversal invariant spin liquids.}
These phases have fully gapped charge degrees of freedom and spin degrees of freedom described by the fermionic spinons ``$f_{i\alpha}$" in \eqref{slave_rotor}.  At the mean-field level, the spinons have an associated band structure that can either be gapped and topological (like a TI) or gapless (like a metal). Thus, the spin sector has an intrinsic single-particle-like nature imposed on it by the formalism.  Relaxing this assumption is one of the ingredients needed to obtain the most exotic fractionalized topological insulators in three dimensions.\cite{Maciejko:prl10,Swingle:prb11,Cho:ap11,Levin:arxiv11}  Indeed, a fruitful approach to studying and/or discovering exotic fractionalized topological insulators in three dimensions may be to deepen our understanding of quantum spin liquids in three dimensions, particularly those whose spinons are strongly correlated and not well described within a single-particle formalism.\cite{Cho:ap11}  In a gapless system, these would be ``algebraic spin liquids".\cite{Balents:nat10}  One could then ``add back" the charge degrees of freedom to obtain an exotic topological insulator.\cite{Kargarian:prb11}  However, it should be emphasized that correspondences between TIs and QSLs described in this paper have all been between QSLs with excitations described by fermions.  At the moment, it remains unclear how to relate spin liquids described by bosonic excitations (Schwinger bosons) to the topological insulators, although some interesting parallels have been noted.\cite{Scharfenberger:prb11}  

\section{Conclusions}
\label{sec:conclusions}

In this work, we have emphasized connections between topological insulators and quantum spin liquids from a variety of viewpoints, with an emphasis on their common topological structure. We have described results from one, two, and three dimensional systems. In one-dimension we have emphasized the important role that quantum entanglement, particularly the entanglement spectrum, can play in studying and classifying phases.  

In two dimensions, we have demonstrated the Kitaev-Haldane-Kane-Mele correspondence and shown how it can play a useful role in finding topological insulators and interesting quantum spin liquids.  In interacting systems we have described the different roles Dirac points and quadratic band crossing points play in the instability to spontaneously generated topological phases, even when there is no intrinsic spin-orbit coupling.  We have given an example of a lattice model of spinless fermions that possesses a very flat band with a finite Chern number.  This model and related ones can be used to help guide the discovery of fractional topological insulators in two dimensions. We have also emphasized transition metal oxide interfaces as a promising experimental system for the realization of interaction-driven topological phases.

In three dimensions, we have focused on semi-realistic models appropriate for transition metal oxides with 4$d$ and 5$d$ electrons.  We have studied the interplay of distortion, interactions, and spin-orbit coupling and presented rich phase diagrams for different $d$-shell fillings.  We have also found a new phase, the ``weak topological Mott insulator", which is expected to reveal itself via topologically protected gapless modes along certain classes of defects.\cite{Ran:np09}  

Looking ahead, it is clear that two of the chief challenges with three-dimensional TIs (which are most likely to appear in applications) are sample quality and reliable tunability.\cite{Fiete:sci11}  To date, the most pressing issue is to obtain samples that have a highly insulating bulk (so that most electrical conductance is along the boundary).  In recent months important advances have been made on this front.  Both Bi$_2$Te$_2$Se\cite{Ren:prb10,Xiong11} and Bi$_{1.5}$Sb$_{0.5}$Te$_{1.7}$Se$_{1.3}$\cite{Taskin11,Ren:prb11} are reported to have more than 70\% of their conductivity on the surface, and there are indications that TI ``work horses" like $\mathrm{Bi_2Se_3}$ are improving in quality, too.\cite{Bansal11}

In the regime of very strong electron interactions (beyond where the slave-rotor formalism should apply), magnetic phases with broken time-reversal symmetry become more likely.\cite{Pesin:np10}  When spin-orbit coupling is strong in this regime, exotic, conventionally ordered phases are also expected to appear,\cite{chen:prb08,Chen:prb10,podolsky:prl09,Podolsky:prb11} including spin-orbital liquid states.\cite{Chen_Balents:prb09,Chen:prl09} It seems likely that the new-found capability to engineer gauge fields in cold-atom systems will play a role in our understanding of TIs and QSLs, too.\cite{Lin:prl09,Spielman:pra09,Lin:nat09,Goldman:cm10,Bermudez:prl10,Stanescu:pra10}  Many novel behaviors certainly await identification, and keeping in mind the ``connections" between various topological phases seems a good strategy to help guide discovery.

\acknowledgments

We gratefully acknowledge helpful discussions with numerous colleagues, including Xiang Hu, Hong Yao, and C.-C. Joseph Wang who collaborated with us on some of the work presented here, and funding from ARO grant W911NF-09-1-0527 and NSF grant DMR-0955778.  GAF is grateful for support from the Banff International Research Station where an abbreviated version of this manuscript was presented in the form of a talk. The authors acknowledge the Texas Advanced Computing Center (TACC) at The University of Texas at Austin for providing computing resources that have contributed to the research results reported within this paper. URL: http://www.tacc.utexas.edu

%

\end{document}